\documentclass[twocolumn]{aastex631}

\received{November 1, 2022}
\revised{December 6, 2022}
\accepted{December 15, 2022}
\submitjournal{ApJ}

\shorttitle{Oscillatory Reconnection as Plasma Diagnostic}

\shortauthors{Karampelas et al.}
\graphicspath{{./}{figures/}}

\begin{document}

\title{Oscillatory reconnection as a plasma diagnostic in the solar corona}

\correspondingauthor{Konstantinos Karampelas}
\email{konstantinos.karampelas@northumbria.ac.uk}

\author[0000-0001-5507-1891]{Konstantinos Karampelas}
\affiliation{Department of Mathematics, Physics and Electrical Engineering, Northumbria University,\\ Newcastle upon Tyne, NE1 8ST, UK}
\affiliation{Centre for mathematical Plasma Astrophysics, Department of Mathematics, KU Leuven,\\ Celestijnenlaan 200B bus 2400, B-3001 Leuven, Belgium }
\author[0000-0002-7863-624X]{James A. McLaughlin}
\affiliation{Department of Mathematics, Physics and Electrical Engineering, Northumbria University,\\ Newcastle upon Tyne, NE1 8ST, UK}
\author[0000-0002-5915-697X]{Gert J. J. Botha}
\affiliation{Department of Mathematics, Physics and Electrical Engineering, Northumbria University,\\ Newcastle upon Tyne, NE1 8ST, UK}
\author[0000-0001-8954-4183]{St\'{e}phane R\'{e}gnier}
\affiliation{Department of Mathematics, Physics and Electrical Engineering, Northumbria University,\\ Newcastle upon Tyne, NE1 8ST, UK}

\begin{abstract}
Oscillatory reconnection is a relaxation process in magnetised plasma, with an inherent periodicity that is exclusively dependent on the properties of the background plasma. This study focuses on the seismological prospects of oscillatory reconnection in the solar corona. We perform three sets of parameter studies (for characteristic coronal values of the background magnetic field, density and temperature) using the PLUTO code to solve the fully compressive, resistive MHD equations for a 2D magnetic X-point. From each parameter study, we derive the period of the  oscillatory reconnection. We find that this period is inversely proportional to the characteristic strength of the background magnetic field and the square root of the initial plasma temperature, while following a square root dependency upon the equilibrium plasma density. These results reveal an inverse proportionality between the magnitude of the Alfv\'en speed and the period, as well as the background sound speed and the period. Furthermore, we note that the addition of anisotropic thermal conduction only leads to a small increase in the mean value for the period. Finally, we establish an empirical formula that gives the value for the period in relation to the background magnetic field, density and temperature. This gives us a quantified relation for oscillatory reconnection, to be used as a plasma diagnostic in the solar corona, opening up the possibility of using oscillatory reconnection for coronal seismology. 
\end{abstract}



\section{Introduction} \label{sec:introduction}

Oscillatory reconnection is a physical phenomenon characterised by a series of reconnection events \citealt{Parker1957JGR,Sweet1958IAUS,Petschek1964NASSP}) that take place alongside periodic changes in the magnetic connectivity of a perturbed magnetic field. The process was identified for the first time in \citet{CraigMcClymont1991ApJ}, during the study of the relaxation of an 2D X-point. One important characteristic of oscillatory reconnection is that the periodicity is not imposed by an external driver, rather it is an inherent property of the relaxation process.

Over the recent years, a number of numerical studies have been conducted regarding oscillatory reconnection. \citet{McLaughlin2009} studied the mechanism for a 2D magnetic X-point in a cold plasma, solving the fully compressible resistive MHD equations. Using an external fast magnetoaccoustic pulse, they initiated oscillatory reconnection by perturbing a magnetic X-point. This study had identified many properties of this mechanism, like the periodic changes in the resulting current sheet orientation with the respective changes in connectivity,  and the formation of both fast and slow oblique magnetic shocks. \citet{Thurgood2017ApJ} later expanded the results of the previous study for a 3D null point, also reporting the generation of MHD waves. Oscillatory reconnection has also been studied for a realistic solar atmosphere, as a result of flux rope emergence \citep{2009A&A...494..329M,McLaughlin2012ApJ}, while other studies revolved around the effects of resistivity, initial perturbation amplitude, and the length of the initial current sheet on the period of the reconnection process \citep{McLaughlin2012A&A, 2018ApJ...855...50T, 2018PhPl...25g2105T, Thurgood2019A&A}. \citet{Stewart2022} reported the onset of oscillatory reconnection and the generation of waves through the coalescence of two cylindrical flux ropes, while \citet{Sabri2020ApJ} reported the development of the plasmoid instability in a magnetic O-point and the resulting manifestation of plasmoid-mediated quasi-oscillatory magnetic reconnection. The results of \citet{McLaughlin2009} have recently been expanded for a hot coronal plasma in \citet{Karampelas2022a}, studying the relation between the oscillation period and the strength of the background magnetic field, while also taking into account the effects of anisotropic thermal conduction. A following study \citep{Karampelas2022b} reported for the first time on the independence between the type and strength of the perturbing wave pulse and the frequency of the resulting oscillatory reconnection in a hot coronal plasma. These two studies have produced encouraging results regarding the possibility of using oscillatory reconnection as a new tool for coronal seismology.

Magnetic reconnection can cause the dissipation of magnetic field and electric current, leading to the acceleration of particles, ejection of mass and heating through the generation of shocks. As such, it is considered as the main mechanism behind solar flares (e.g. \citealt{ShibataMagara2011LRSP,2015ApJ...812..105J}), while the ubiquitous null points in the solar atmosphere \citep{Galsgaard1997,BrownPriest2001AnA,Longcope2005LRSP,Regnier2008AnA}, where reconnection can take place, are consequently considered preferential locations the manifestation of flares (e.g. \citealt{2011A&A...533A..18M}). Over the years, oscillatory reconnection has been proposed as a driving force behind observed phenomena like quasi-periodic pulsations (QPPs) of solar flares \citealt{Kupriyanova2016SoPh,VanDoorsselaere2016SoPh,Pugh2017AnA,2019ApJ...886L..25Y,2020ApJ...895...50H, 2020A&A...639L...5L,2020ApJ...893....7L,2021ApJ...921..179L, 2022FrASS...932099L,   2021ApJ...910..123C, 2022ApJ...931L..28L, 2022RAA....22j5017S}) and stellar flares (e.g. \citealt{2019A&A...629A.147B,2019A&A...622A.210G,2019MNRAS.482.5553J,2019ApJ...876...58N,2019ApJ...884..160V,2020A&A...636A..96M, 2021SoPh..296..162R}). The mechanism is included in reviews summarising our current knowledge around QPPs and the proposed mechanisms behind them, such as  \citet{McLaughlin2018SSRv}, \citet{Kupriyanova2020STP}, and \citet{Zimovets2021SSRv}. In particular, there are many examples from QPP observations  \citep[see histogram in][and its corresponding online catalog]{McLaughlin2018SSRv}, with reported periods close to those derived from the studies of \citet{Karampelas2022a} and \citet{Karampelas2022b}, for the plasma conditions considered in those studies.

Connection has also been proposed between oscillatory reconnection and quasi-periodic flows, like those associated with spicules (e.g. \citealt{DePontieu2010ApJ,DePontieu2011Sci,2019Sci...366..890S, 2020ApJ...891L..21Y}), as well as with observed periodicities in breakout current sheets at the base of jets \citep{2019ApJ...874..146H}.  \citet{Mandal2022a}, have reported a highly dynamic small-scale jet in a polar coronal hole, and proposed oscillatory reconnection as a possible driving mechanism behind the observed repetitive outflows. \citet{McLaughlin2012ApJ} were able to reproduce such observed periodic outflows through oscillatory reconnection in a 2D flux emergence model. The resulting periods from that model had a very good match with those reported from wavelet analysis in \citet{Mandal2022a}, although the latter showed no significant power at the $99 \%$ confidence level, preventing them to characterise the outflows as periodic, but merely repetitive. Observational signatures of chromospheric jets by periodic reconnection events were also reported in simulations by \citet{Heggland2009ApJ}, although, the periodicity was attributed to the continuous driving rather than being inherent to the system. Oscillatory reconnection has also been considered as a possible mechanism behind with the creation of an observed quasi-periodic fast-propagating (QFP) magnetosonic wave from the eruption of a magnetic flux rope \citep{2018ApJ...853....1S}, as well as behind the formation and disappearance of a small scale magnetic flux rope consisting of new loops formed by the reconnection events \citep{2019ApJ...874L..27X}. \citet{Zhang2014A&A} have reported oscillatory (or reciprocatory) magnetic reconnection in observations of Coronal Bright Points (CBPs), while reversals of an elongated current sheet in a recent numerical 2D CBP model has been attributed to oscillatory reconnection \citep{NobregaSiverio2022ApJ}. Finally, recent observations by the Parker Solar Probe could also be attributed to oscillatory reconnection (e.g. \citealt{2016SSRv..204...49B, 2019Natur.576..237B,2019Natur.576..228K}), like Alfv\'enic spikes \citep{2021ApJ...913L..14H} and periodicities correlated with Type III radio bursts \citep{2021A&A...650A...6C}.

In this paper, we will further investigate oscillatory reconnection in a hot coronal plasma and to explore its potential for utilising  oscillatory reconnection as a tool for coronal seismology. We will expand the results of \citet{Karampelas2022a} through a series of parameter studies for different characteristic strengths of the magnetic field, equilibrium plasma density and initial plasma temperature, for a 2D magnetic X-point. Like in \citet{Karampelas2022a} and \citet{Karampelas2022b}, we will be exploring these cases both in the absence and presence of anisotropic thermal conduction. In section \ref{sec:setup} we present our physical domain, code used to solve the fully compressible mhd equations and numerical schemes utilised, while we present the results of the parameter studies in the respective subsection in \S\ref{sec:results}. Finally, our conclusions and general discussion take place in \S\ref{sec:discussions}.


\section{Numerical setup} \label{sec:setup}

\begin{figure}[t]
    \centering
    \includegraphics[trim={1.cm 0.5cm 1.cm 0.cm},clip,scale=0.45]{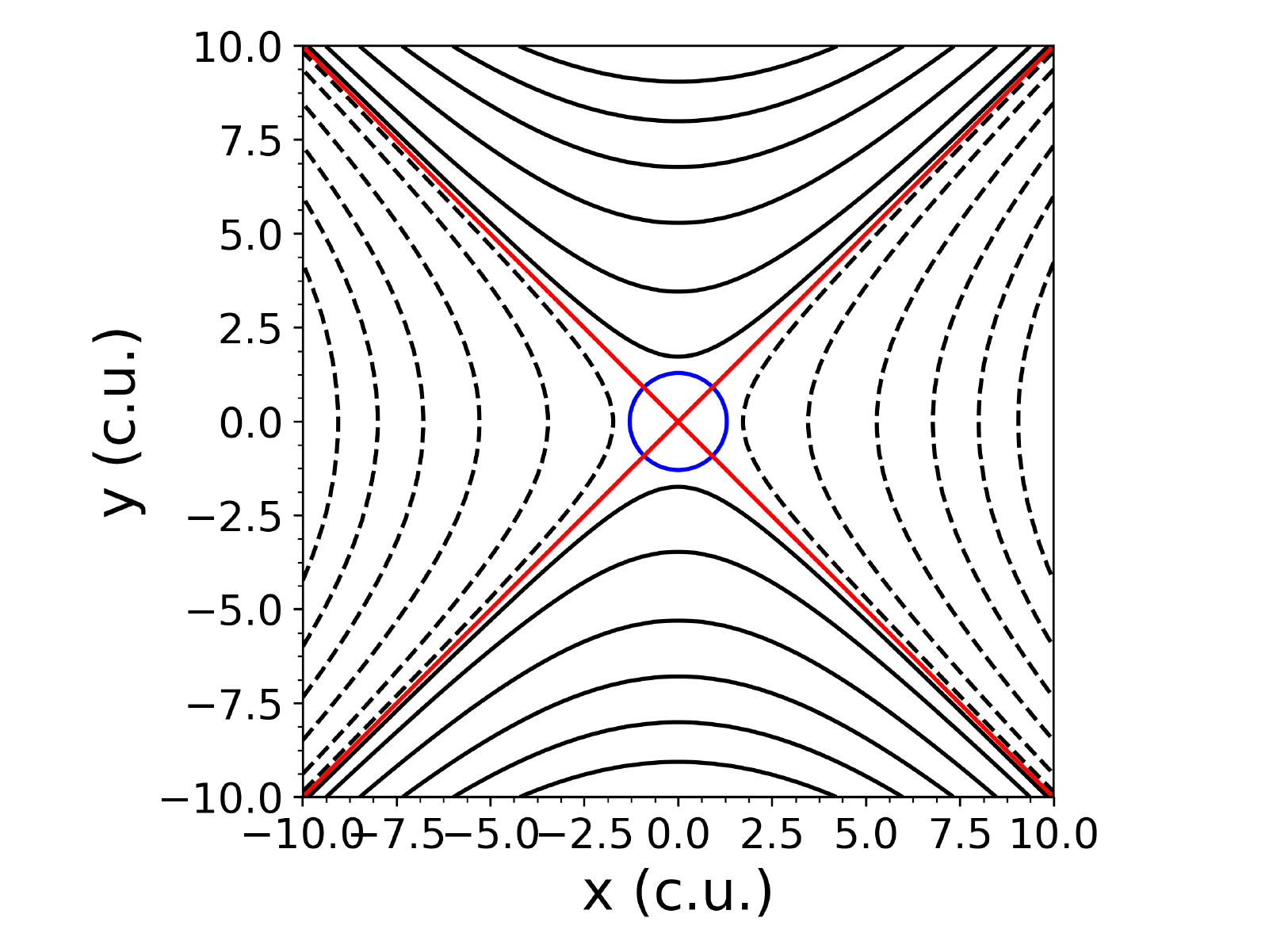}
    \caption{Magnetic field lines of the unperturbed X-point, were the black solid and dashed lines depict the regions of opposing polarity. The separatrices (red solid lines) and the equipartition layer for a $1$\,MK coronal plasma (blue circular line) are also included.}
    \label{fig:profileB}
\end{figure}

\begin{figure}[t]
    \centering
    \resizebox{\hsize}{!}{
    \includegraphics[trim={0.cm  0cm 0.cm 0.cm},clip,scale=0.45]{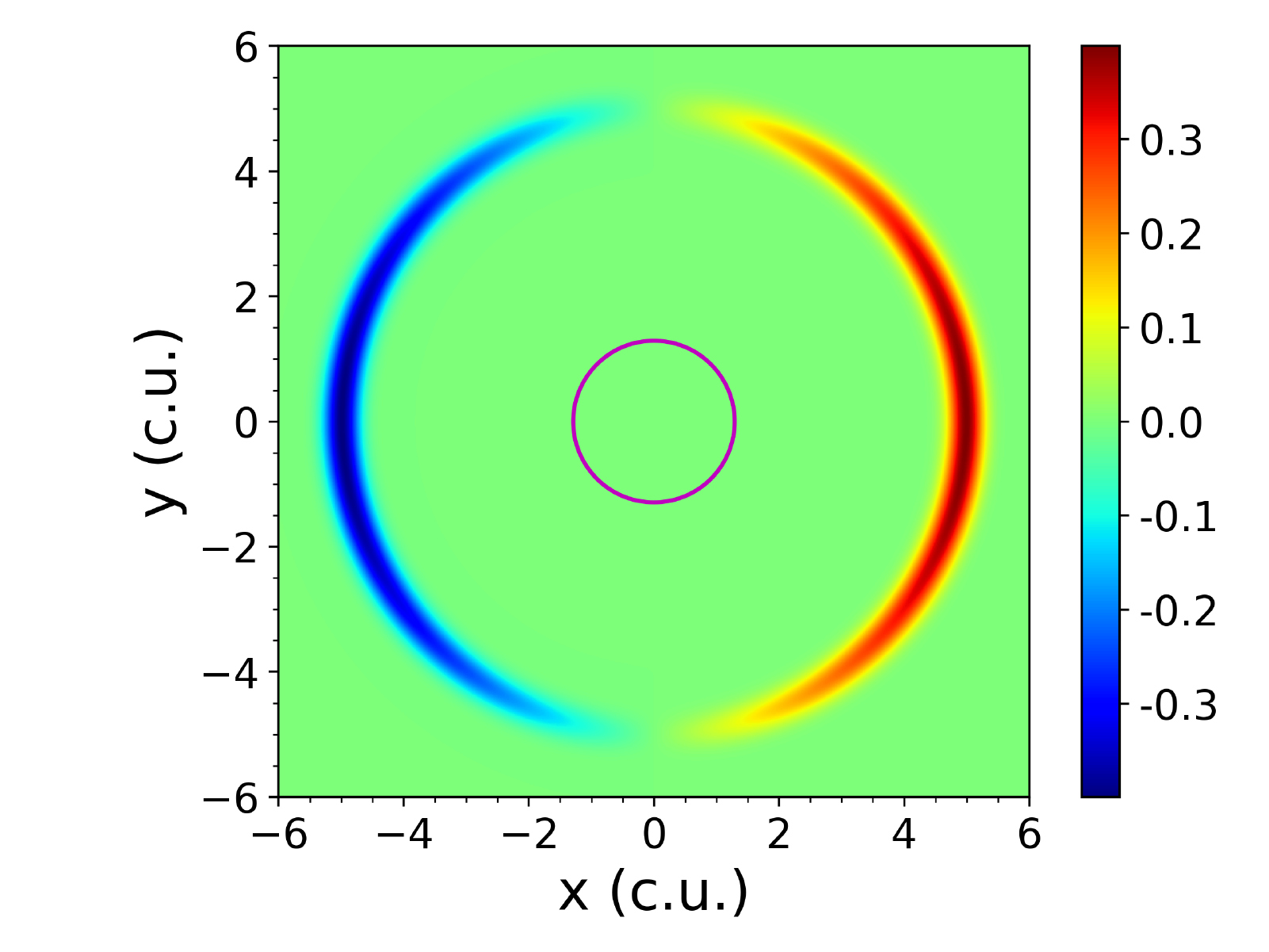}
    \includegraphics[trim={3.cm  0cm 0.cm 0.cm},clip,scale=0.45]{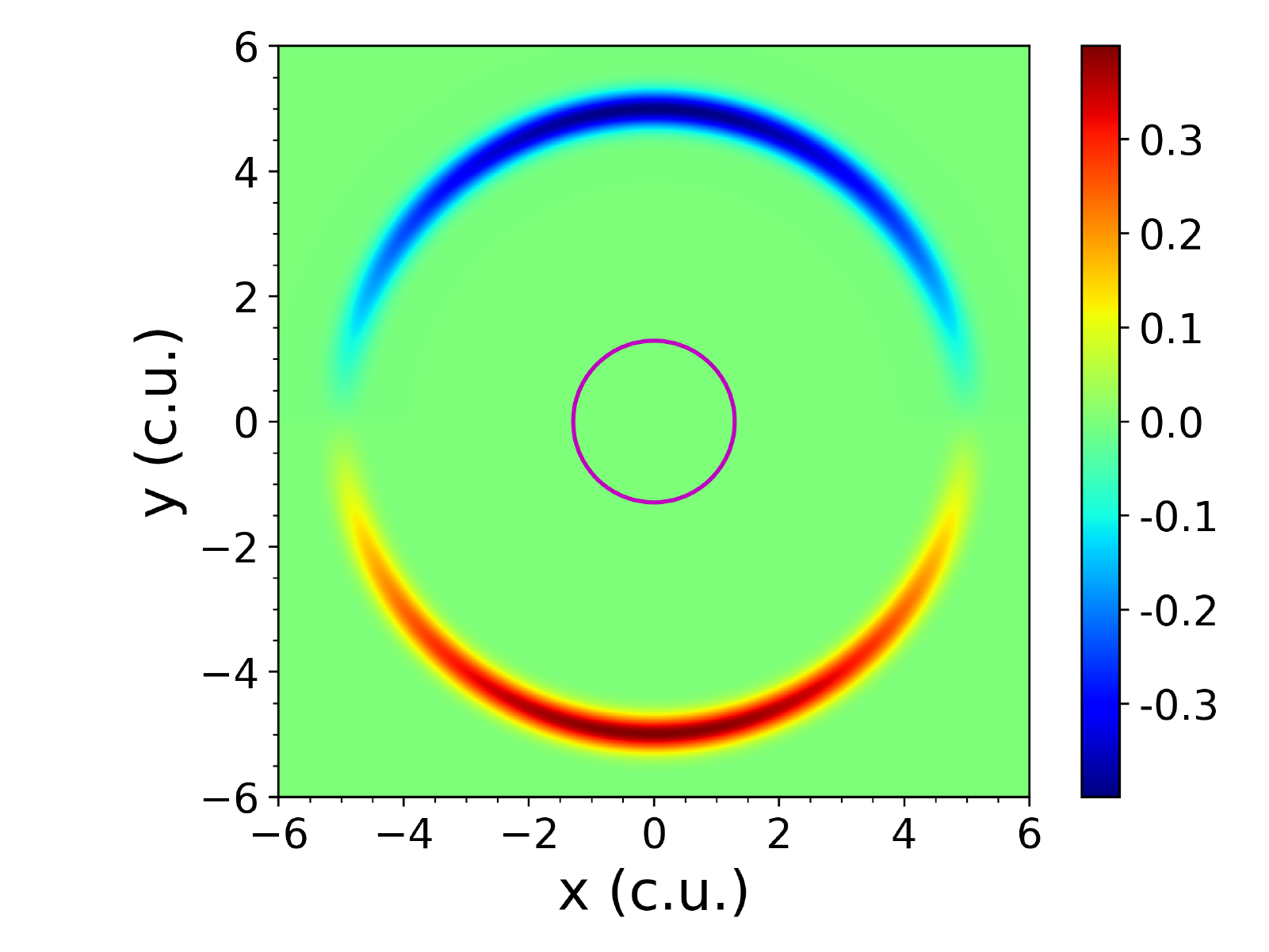}
    }
    \caption{2D profiles of the $v_x$ and $v_y$ velocity components of the initial circular pulse. The magenta circular line is the equipartition layer for a $1$\,MK coronal plasma. All values are depicted in code units.}
    \label{fig:profiledriver}
\end{figure}

\begin{figure*}[t]
    \centering
    \resizebox{\hsize}{!}{
    \includegraphics[trim={1.cm 1.08cm 0.cm 0.cm},clip,scale=0.4]{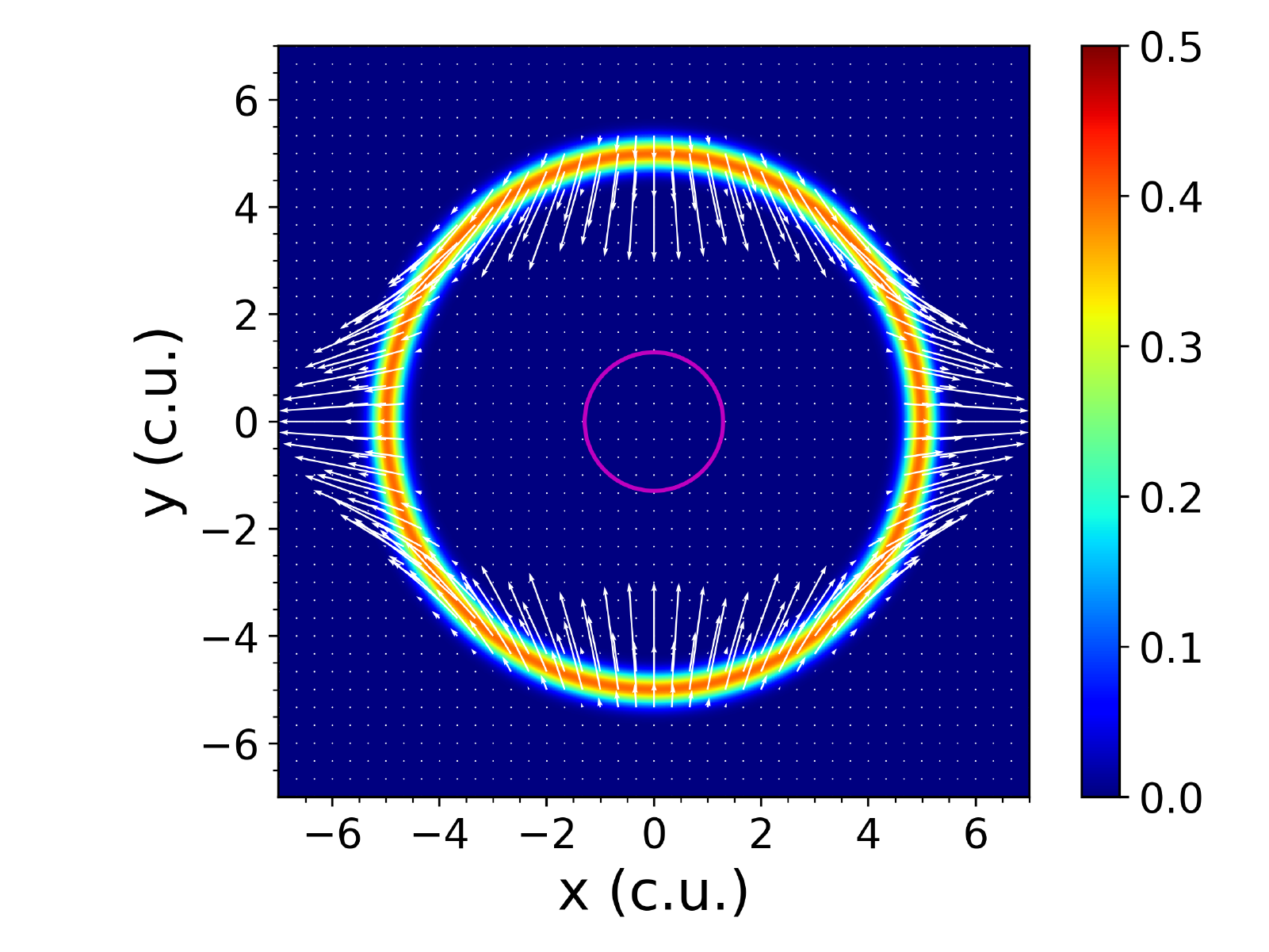}
    \includegraphics[trim={3.cm 1.08cm 0.cm 0.cm},clip,scale=0.4]{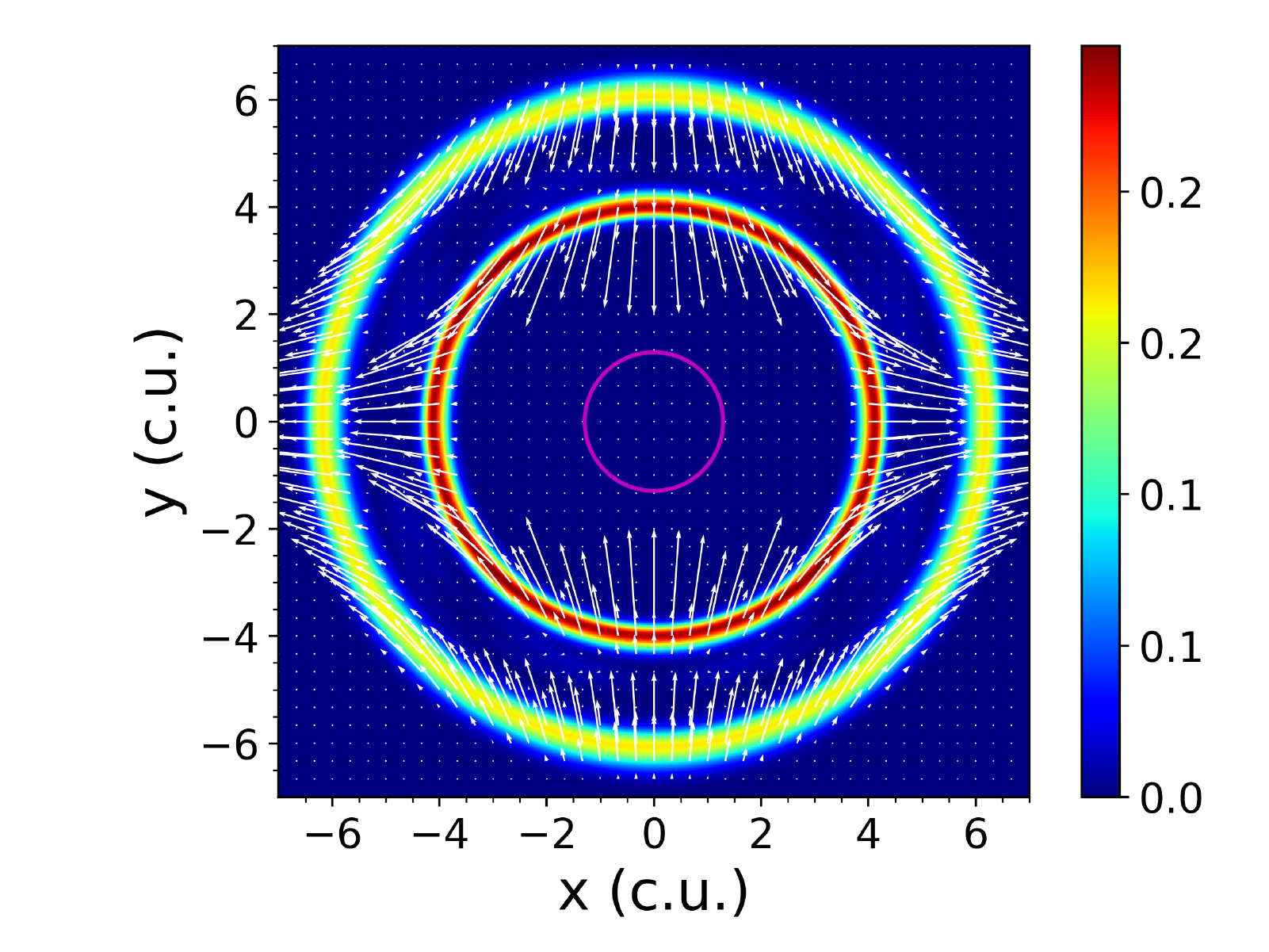}
    \includegraphics[trim={3.cm 1.08cm 0.cm 0.cm},clip,scale=0.4]{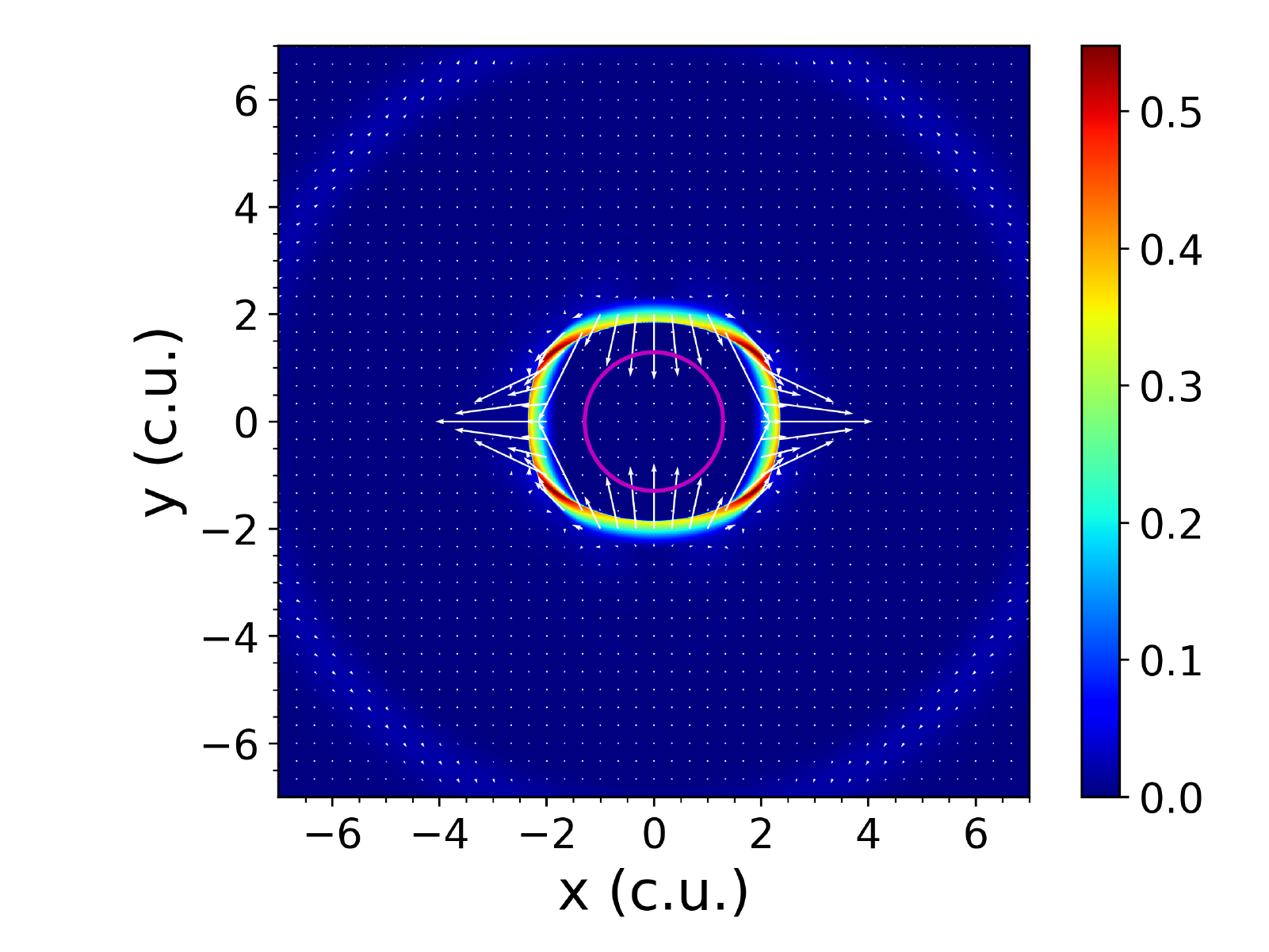}
    }
    \resizebox{\hsize}{!}{
    \includegraphics[trim={1.cm 0cm 0.cm 0.cm},clip,scale=0.4]{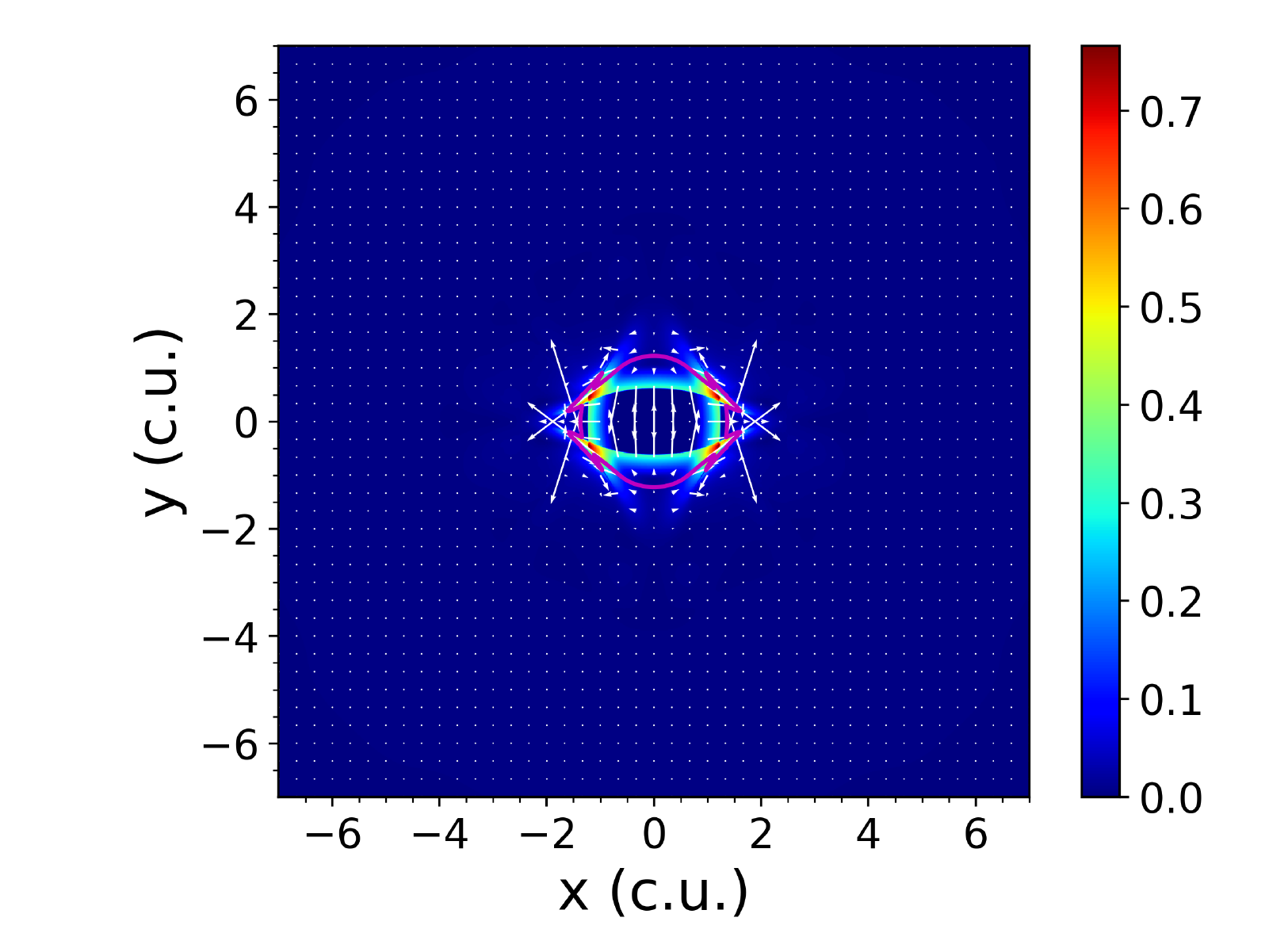}
    \includegraphics[trim={3.cm 0cm 0.cm 0.cm},clip,scale=0.4]{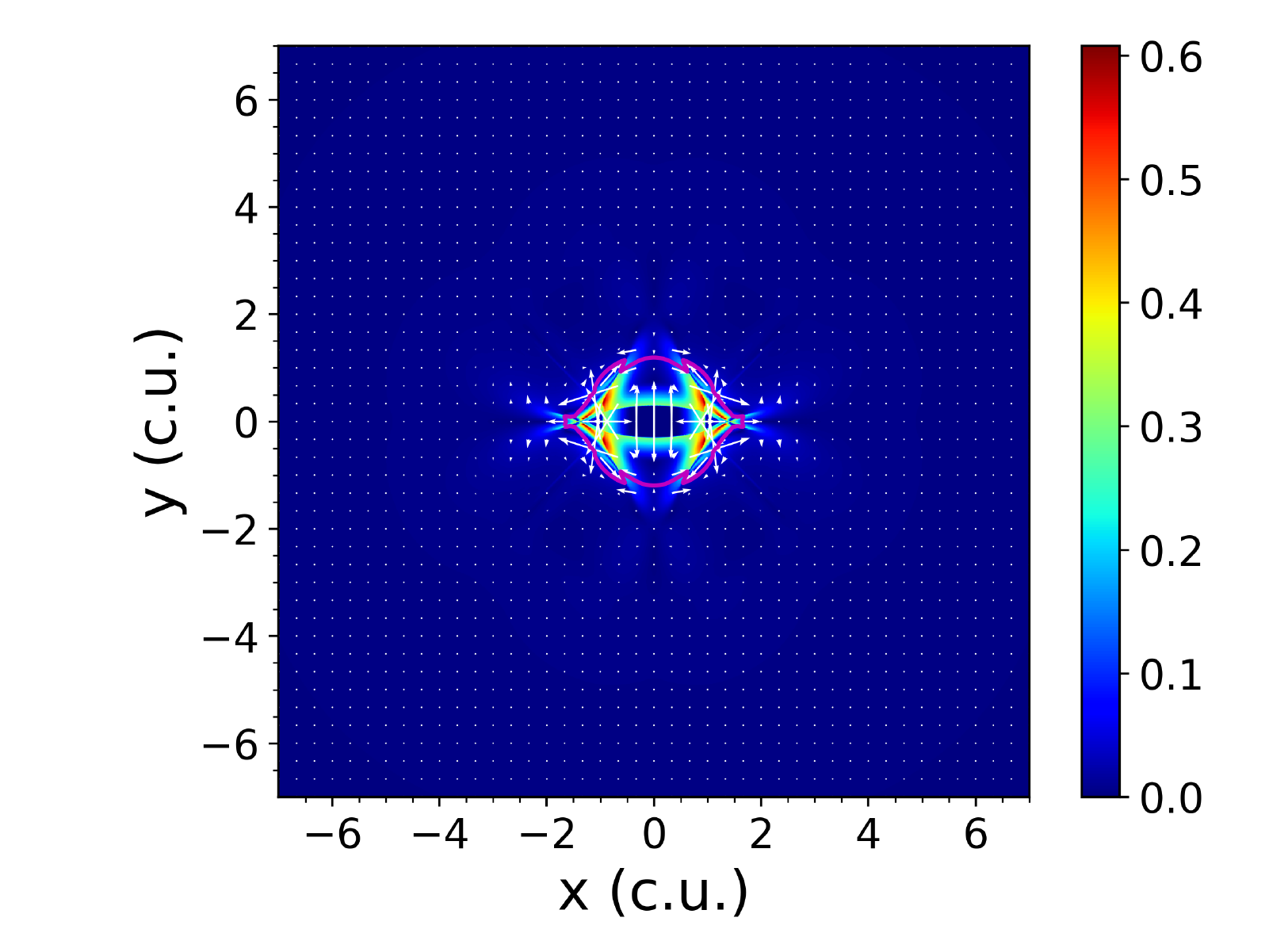}
    \includegraphics[trim={3.cm 0cm 0.cm 0.cm},clip,scale=0.4]{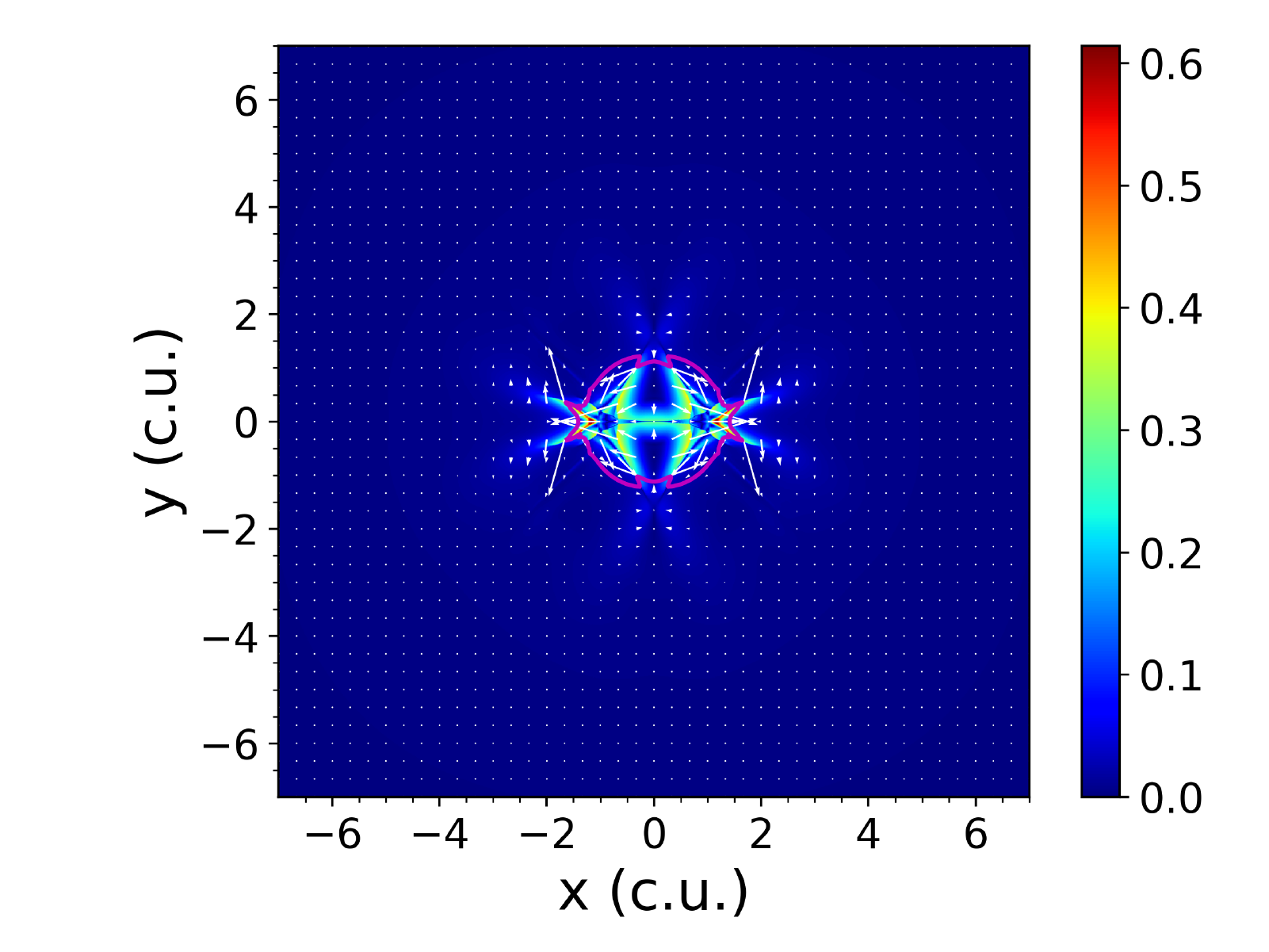}
    }
    \caption{The evolution of the absolute value of the radial velocity for Model 2 (see Table \ref{tab:param}), and the respective vector plot (normalized). Starting from the top, from left to right, the snapshots correspond to time $t=0,\, 0.2,\, 0.8,\, 1.4,\, 1.6$ and $1.8\,t_0$.  All values are depicted in code units.}
    \label{fig:pulevol}
\end{figure*}

\begin{figure*}[t]
    \centering
    \includegraphics[trim={0.4cm 0.cm 1.4cm 0.8cm},clip,scale=0.45]{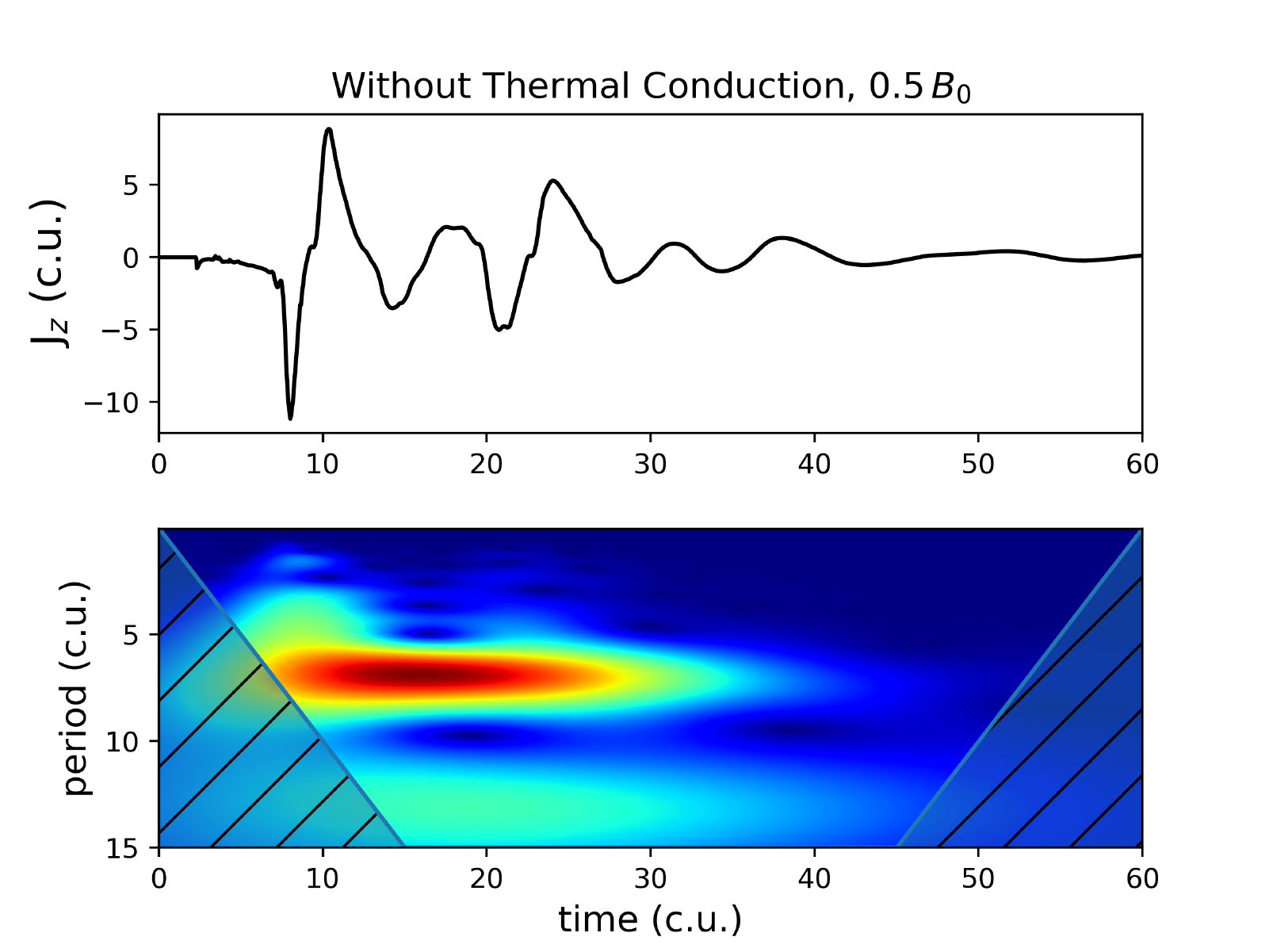}
    \includegraphics[trim={0.4cm 0.cm 1.4cm 0.8cm},clip,scale=0.45]{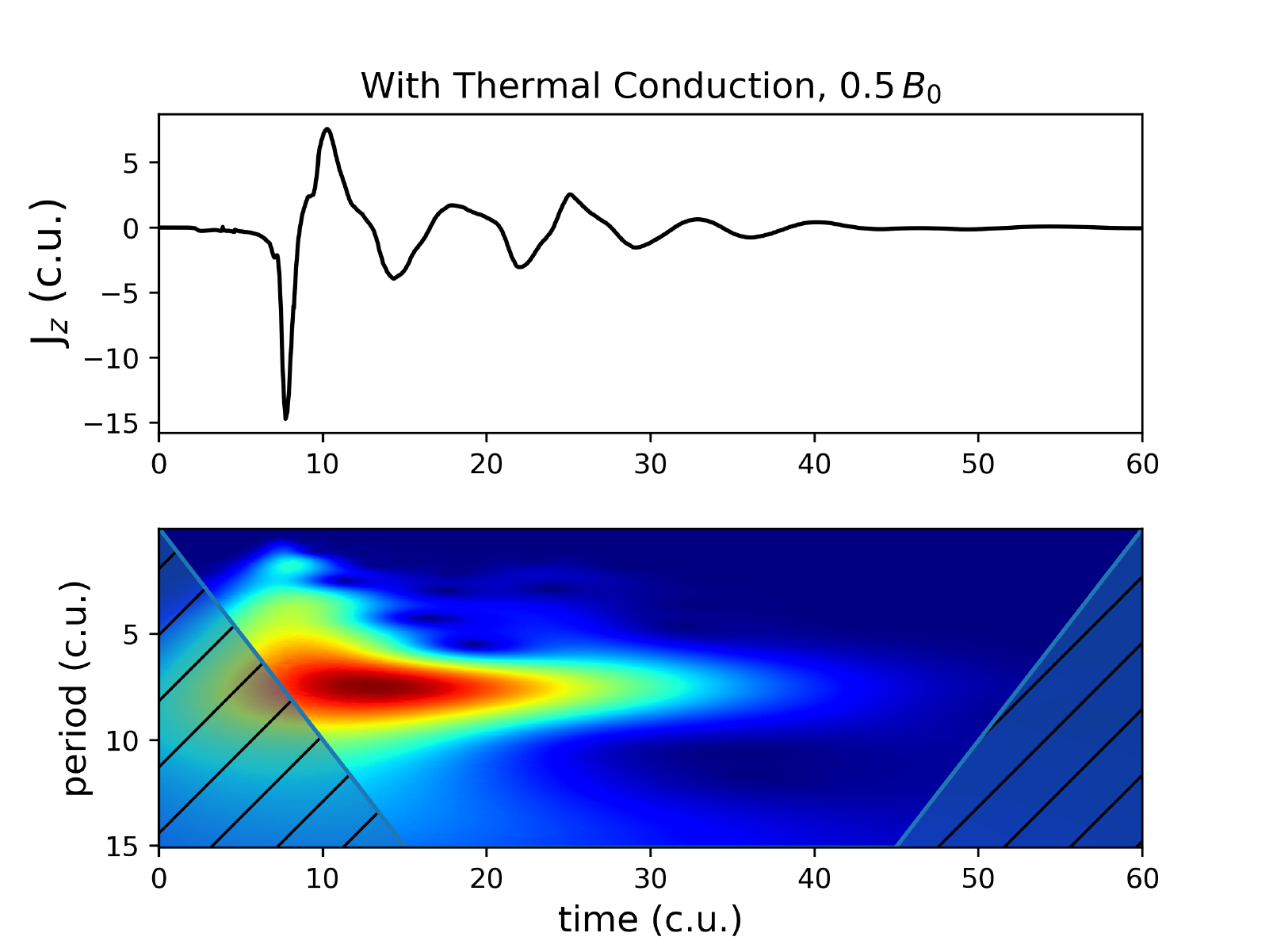}
    
    \includegraphics[trim={0.4cm 0.cm 1.4cm 0.8cm},clip,scale=0.45]{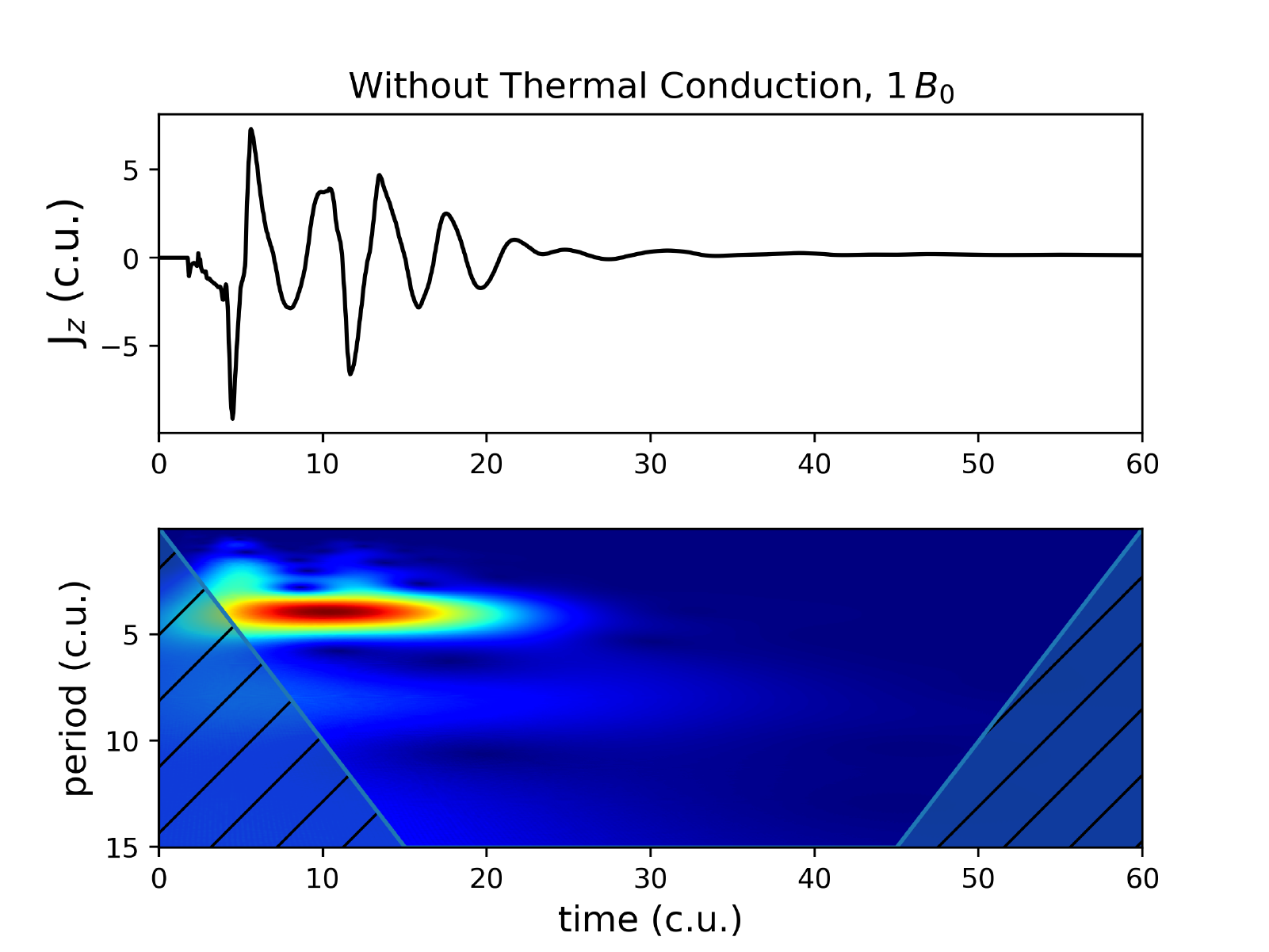}
    \includegraphics[trim={0.4cm 0.cm 1.4cm 0.8cm},clip,scale=0.45]{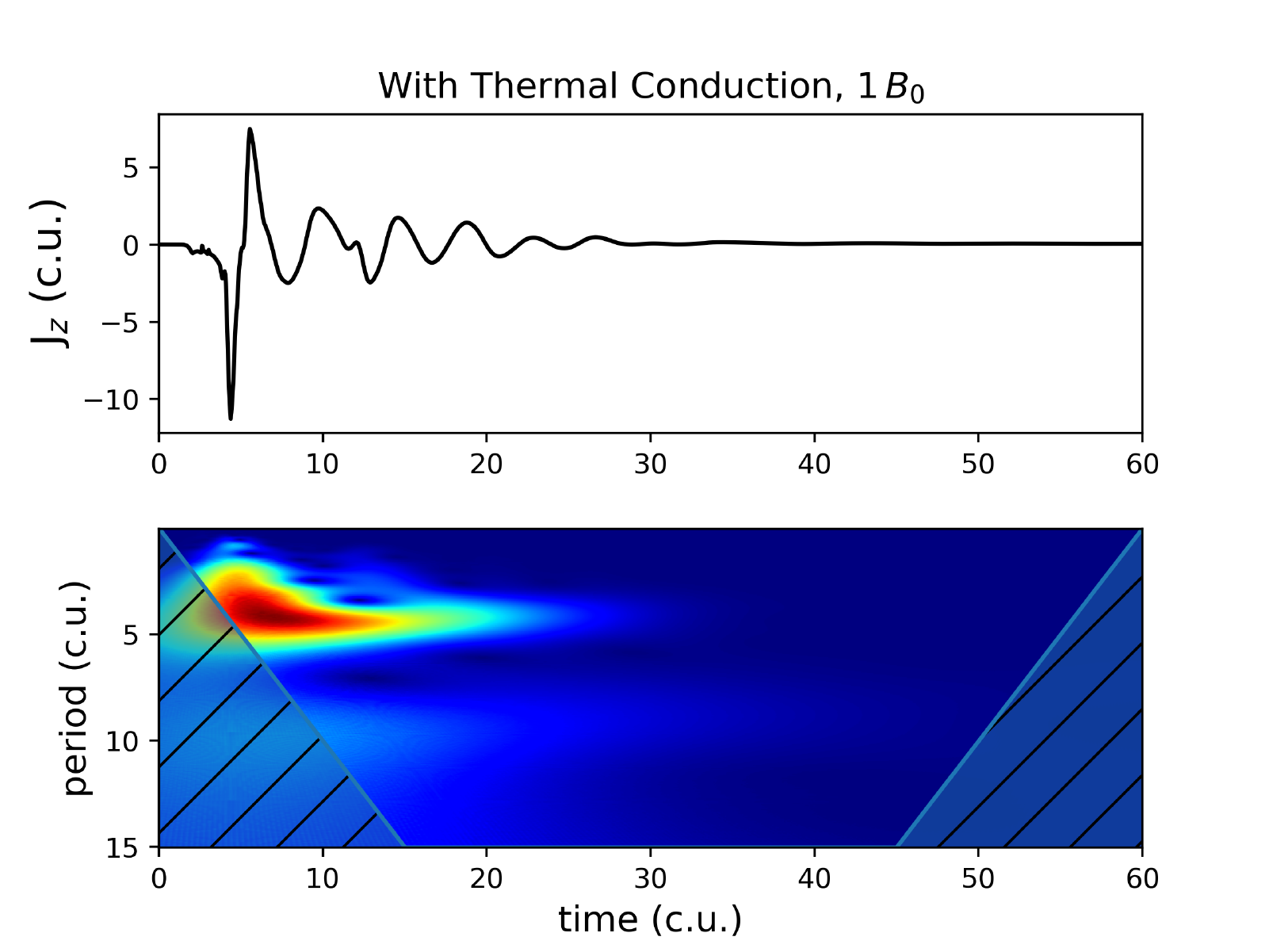}
    
    \includegraphics[trim={0.4cm 0.cm 1.4cm 0.8cm},clip,scale=0.45]{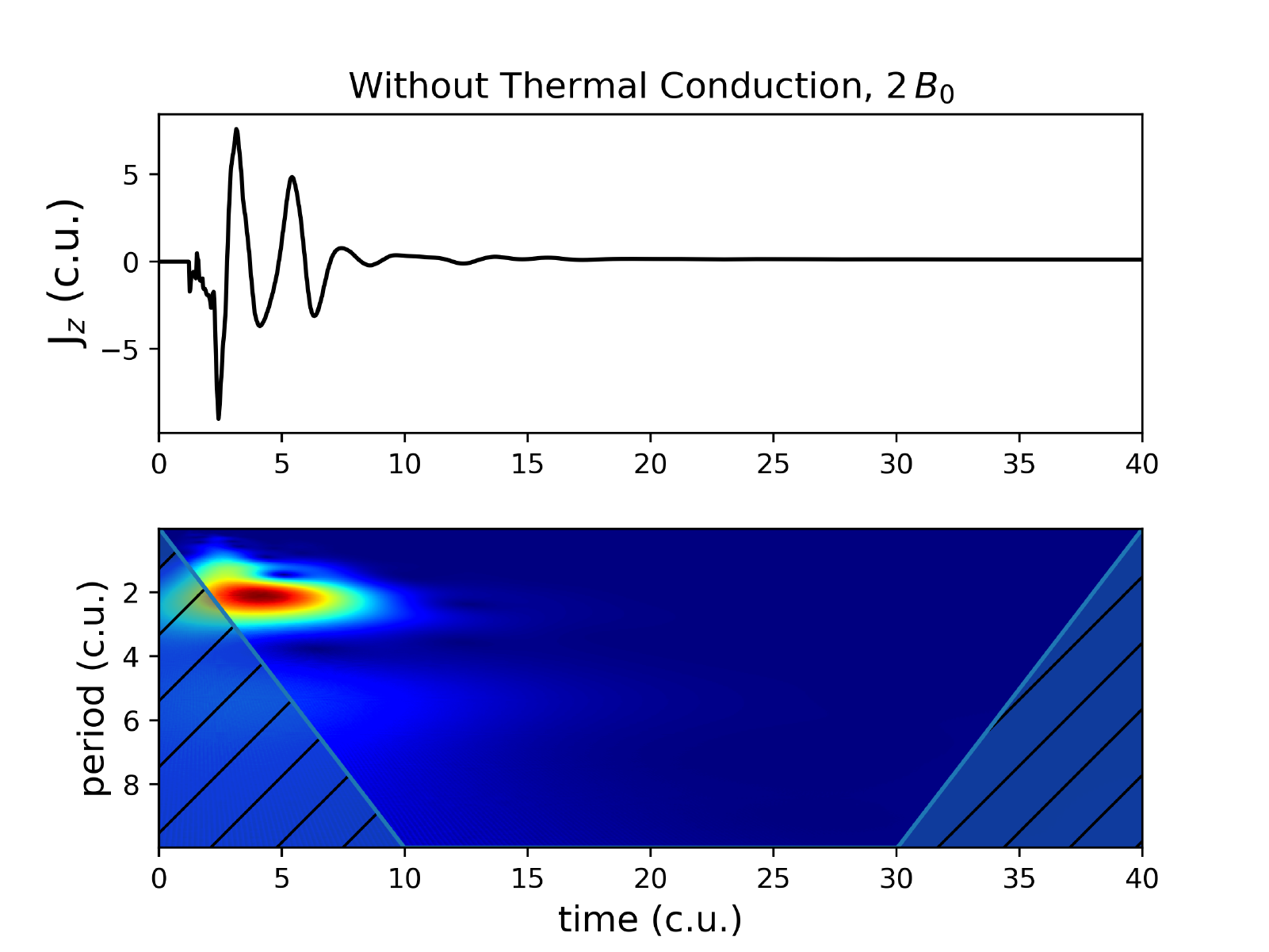}
    \includegraphics[trim={0.4cm 0.cm 1.4cm 0.8cm},clip,scale=0.45]{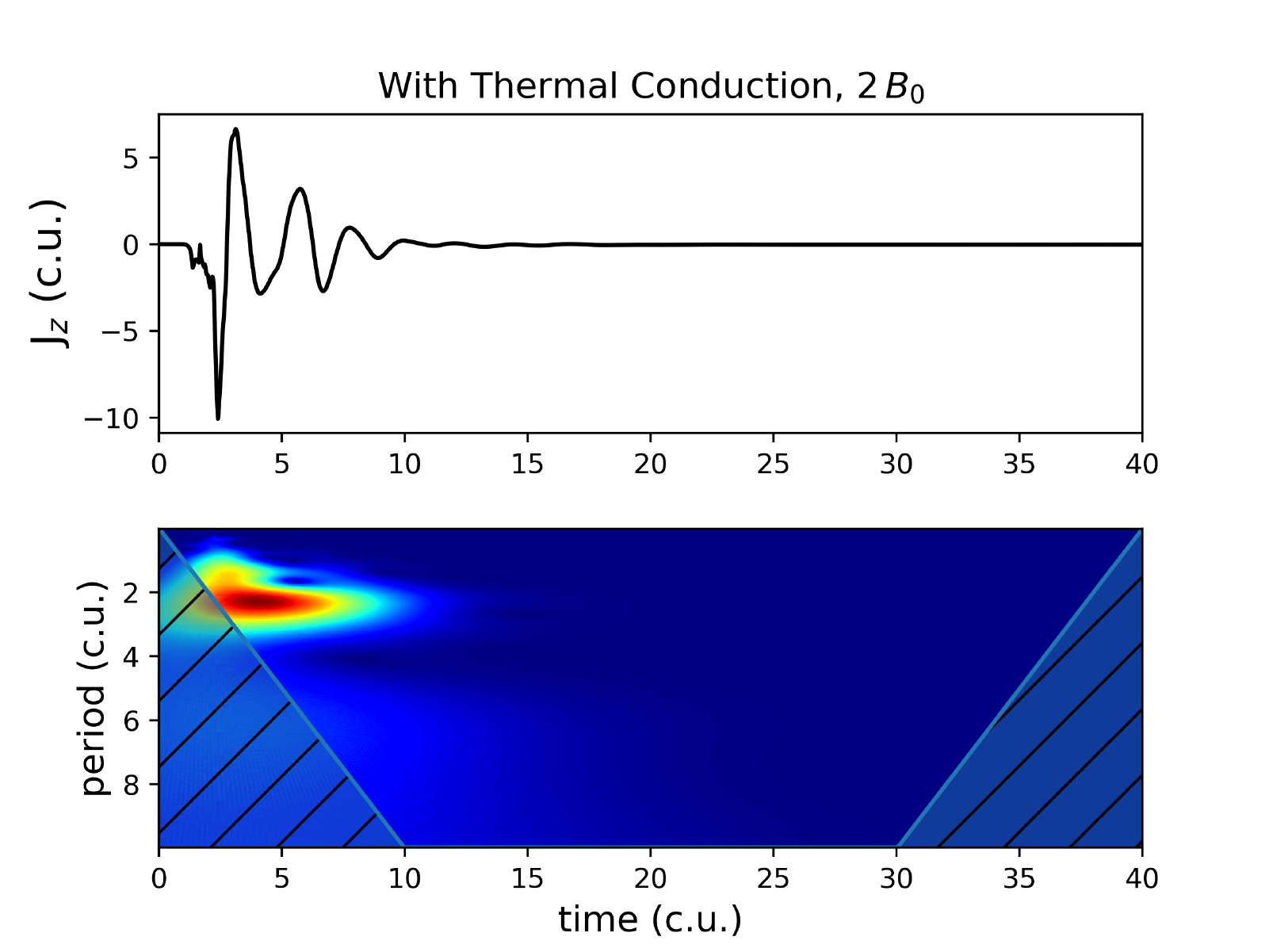}
    \includegraphics[trim={0.4cm 0.cm 1.4cm 0.8cm},clip,scale=0.45]{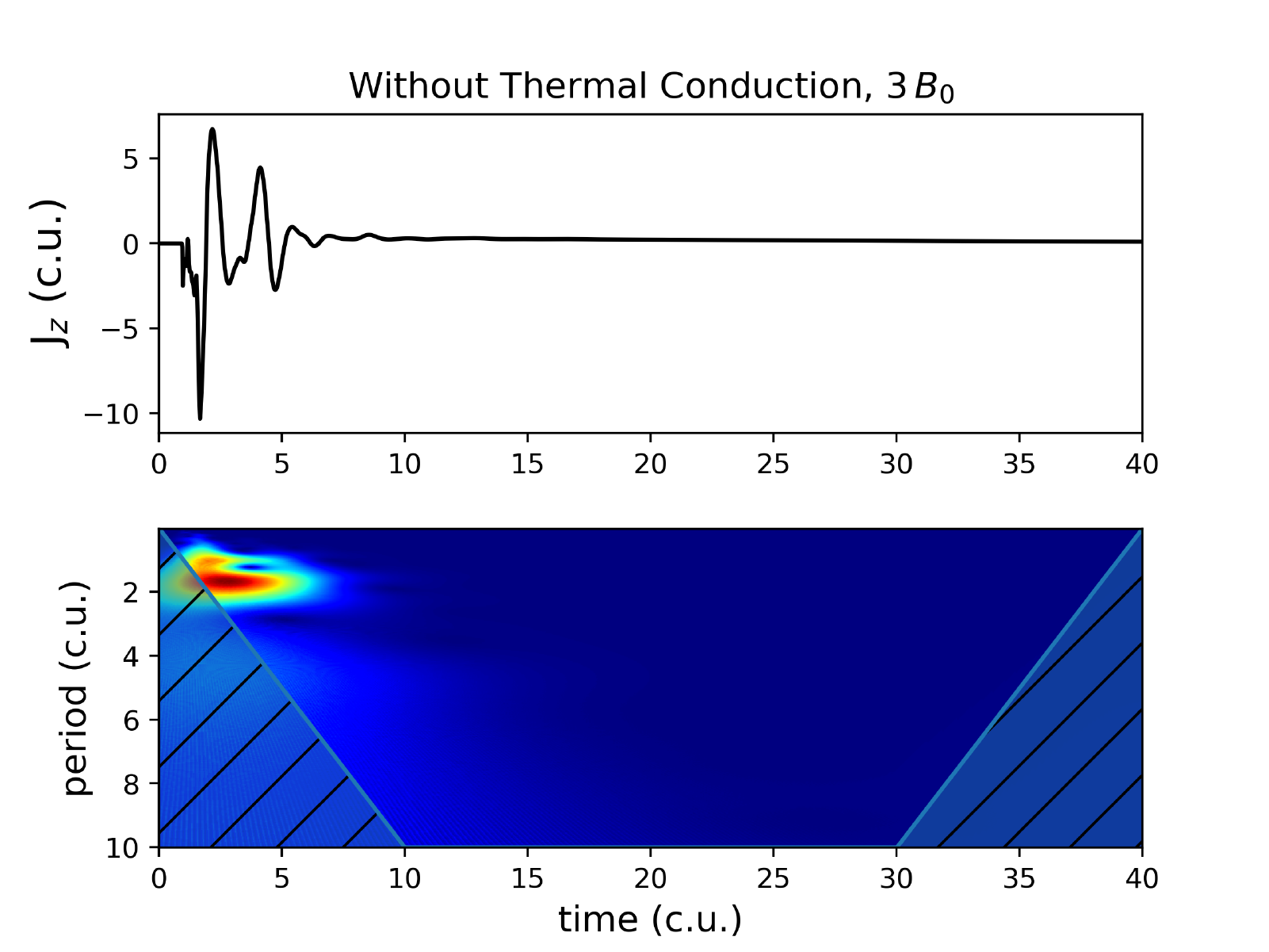}
    \includegraphics[trim={0.4cm 0.cm 1.4cm 0.8cm},clip,scale=0.45]{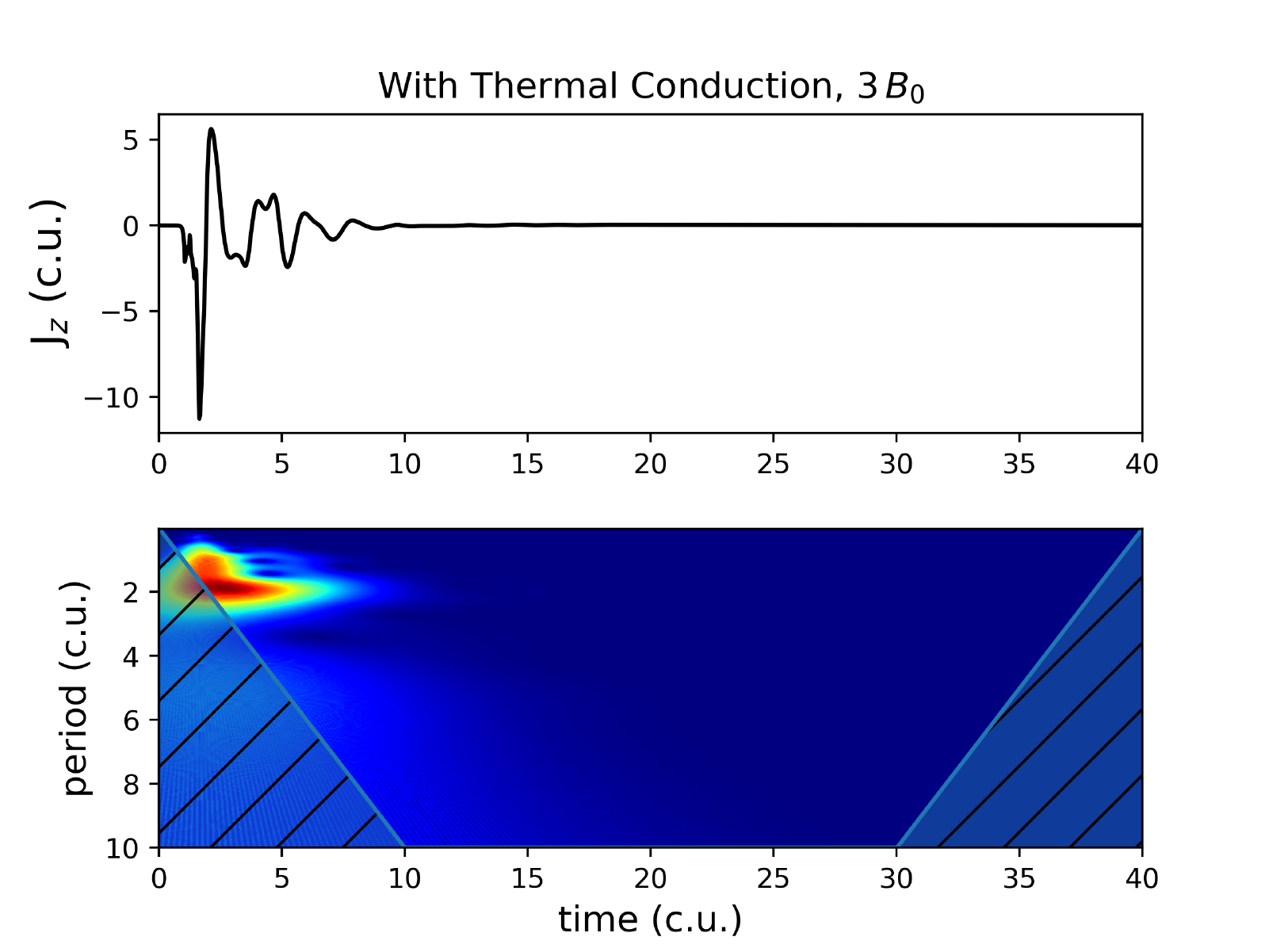}
    \caption{Time series of the $J_z$ current density at the null point for setups with different characteristic magnetic field strength ($0.5\,B_0,\,1\,B_0,\,2\,B_0,$ and $\,3\,B_0$). The equilibrium density and temperature are $1\,\rho_0$ and $1$\,MK respectively. Cases without (left column) and with anisotropic thermal conduction (right column) are considered. All values are depicted in code units.}
    \label{fig:waveletB}
\end{figure*}

\begin{figure}[t]
    \centering
    \includegraphics[trim={0.cm 0.cm 0.cm 0.cm},clip,scale=0.45]{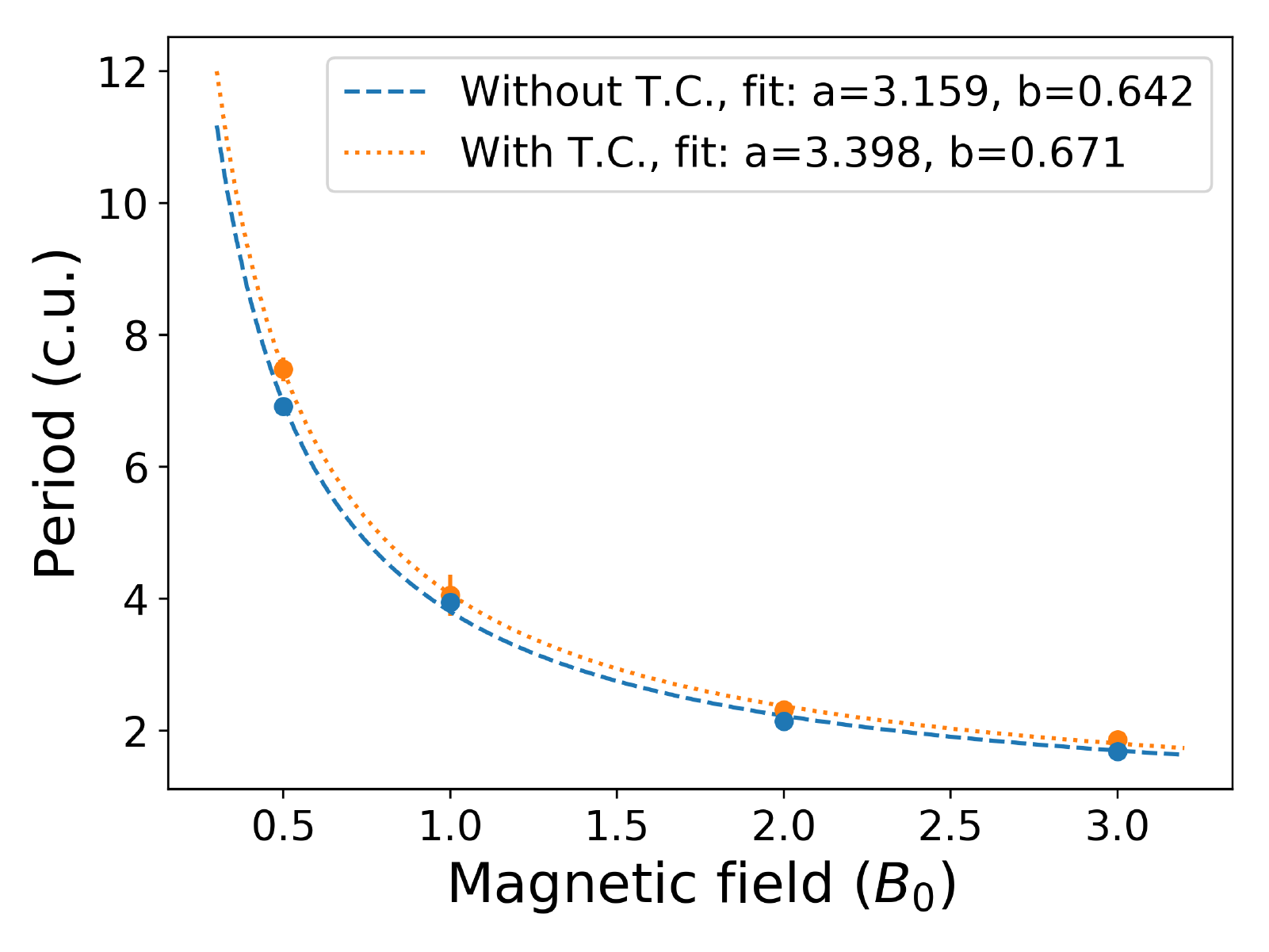}
    \caption{Graph depicting the distribution of the $J_z$ oscillation period with respect to the magnetic field magnitude at radius $r=1$. Overplotted are the fits for both distributions of the function $F(B_0)=a\,(B_0)^{-1}+b$. The blue dashed and orange dotted lines correspond to the cases without and with anisotropic thermal conduction. All values are depicted in code units, unless stated otherwise.}
    \label{fig:PerMagn}
\end{figure}

\begin{figure*}[t]
    \centering
    \includegraphics[trim={0.4cm 0.cm 1.4cm 0.8cm},clip,scale=0.45]{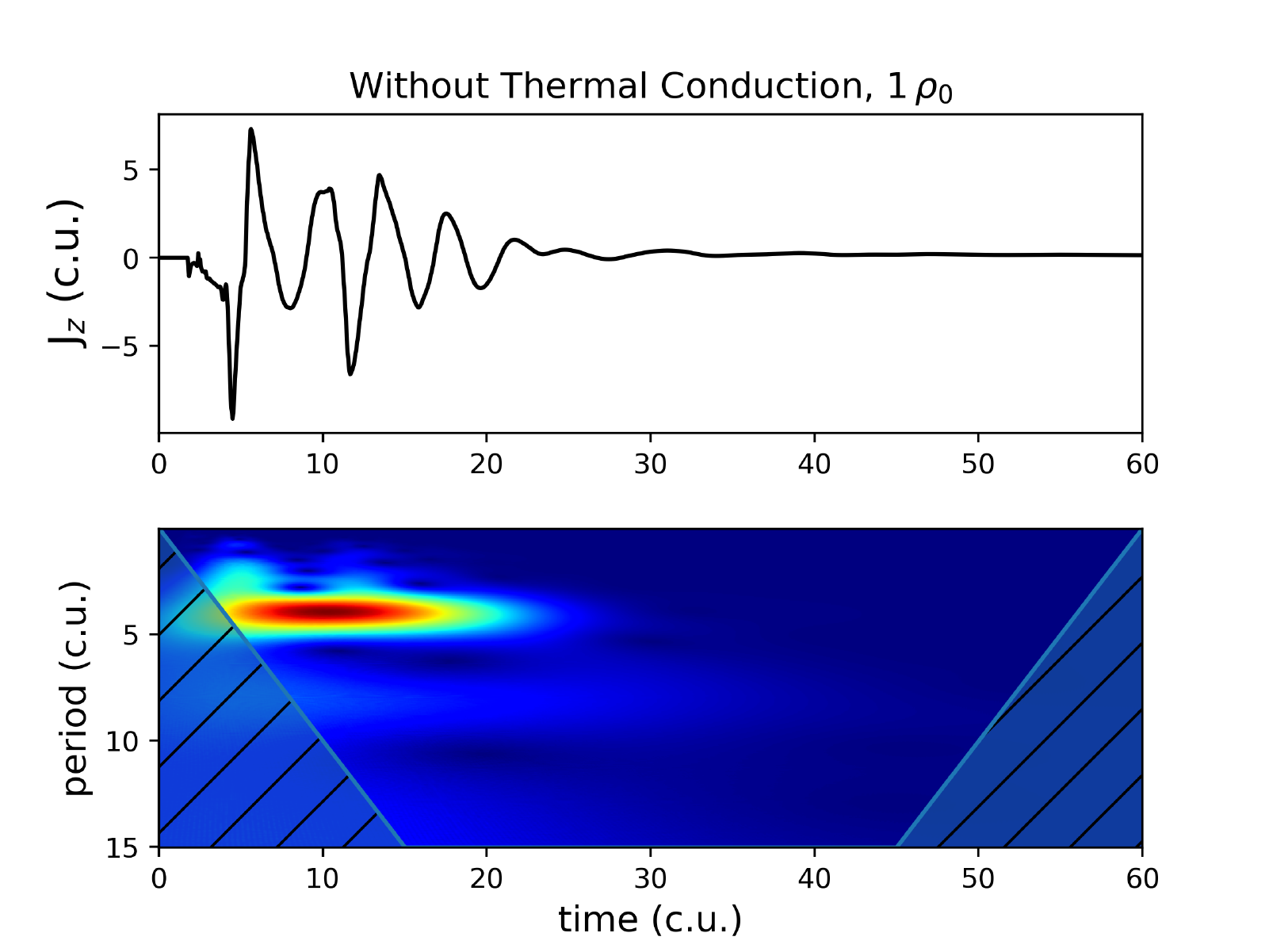}
    \includegraphics[trim={0.4cm 0.cm 1.4cm 0.8cm},clip,scale=0.45]{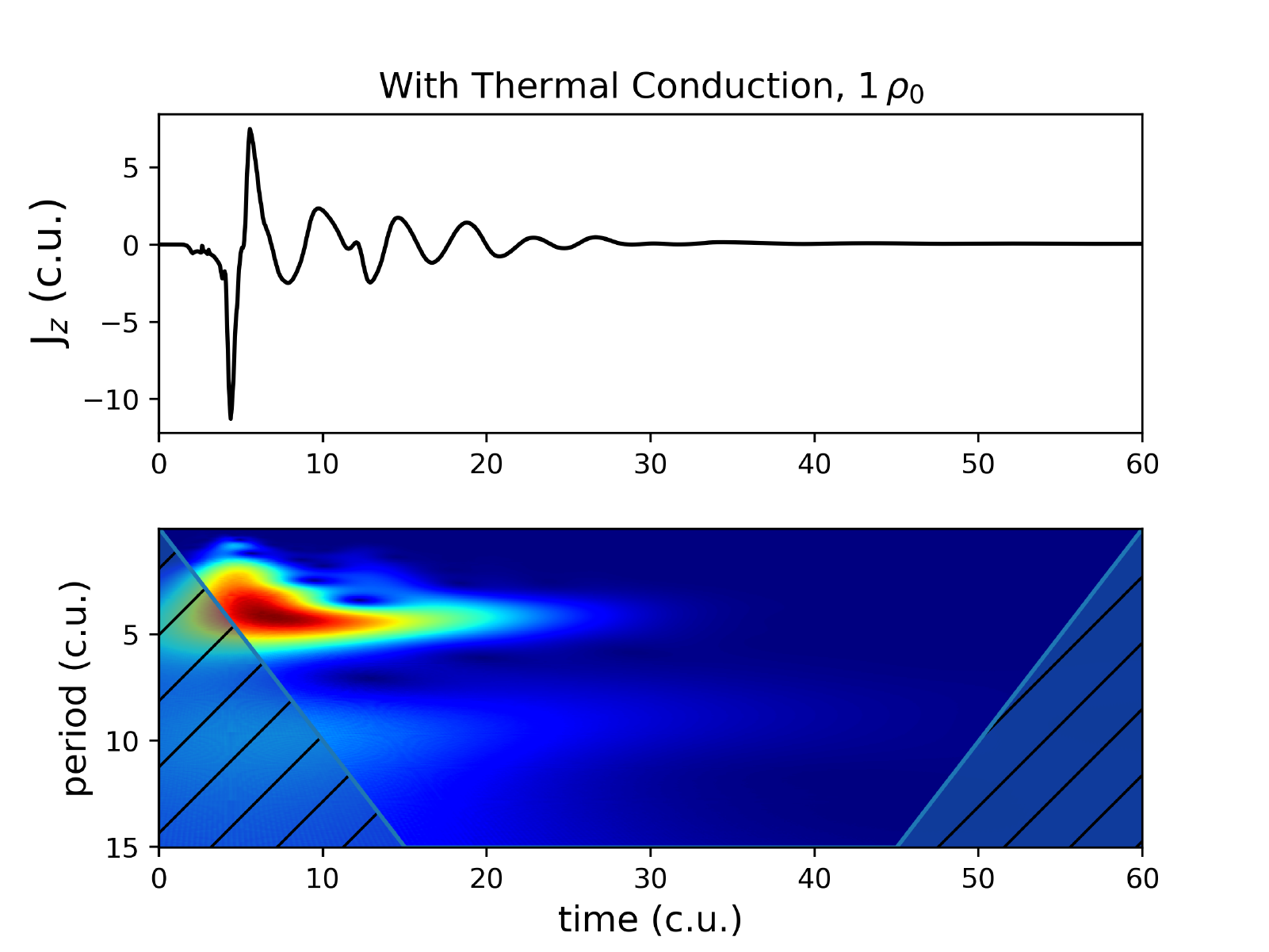}
    \includegraphics[trim={0.4cm 0.cm 1.4cm 0.8cm},clip,scale=0.45]{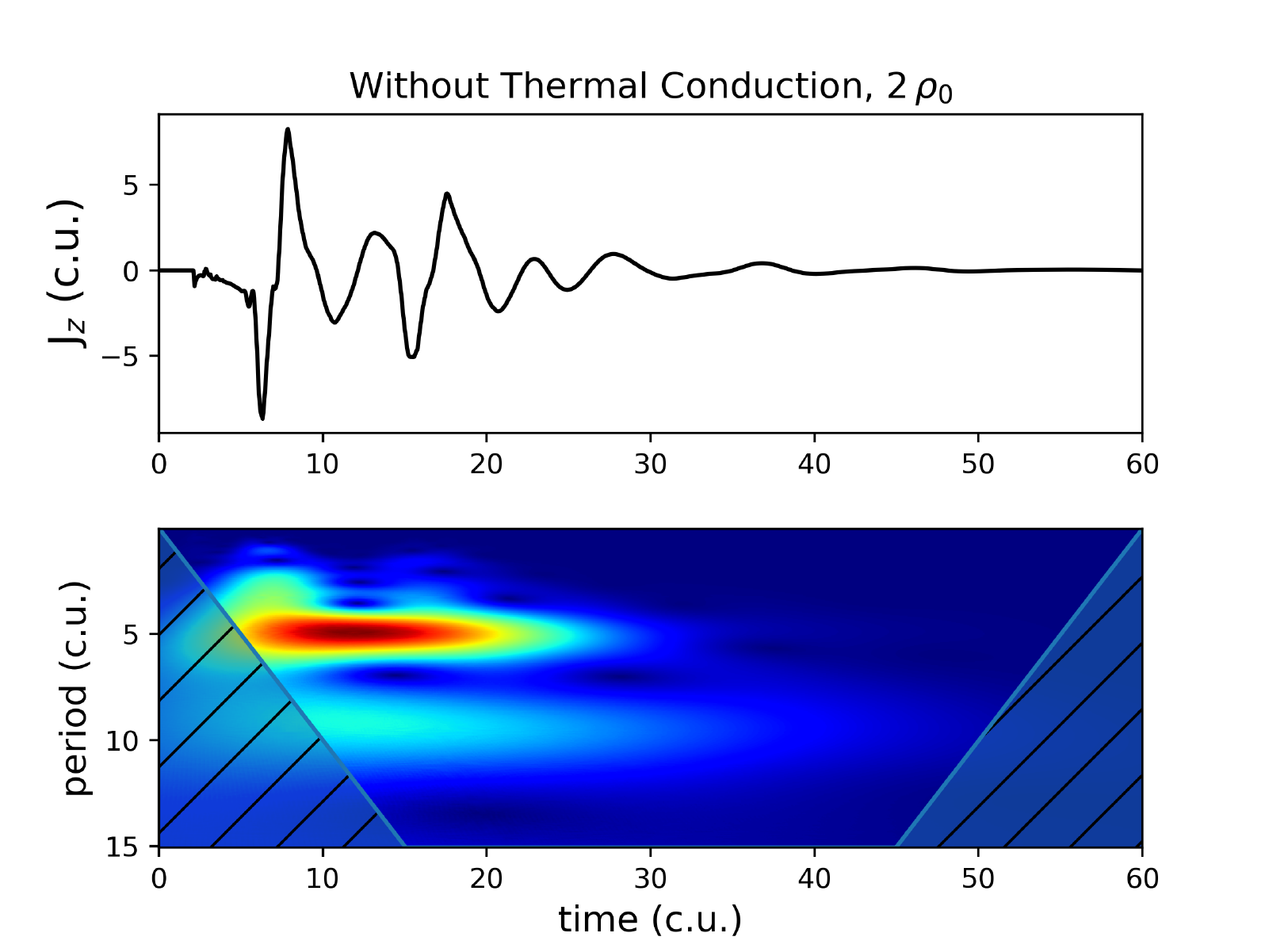}
    \includegraphics[trim={0.4cm 0.cm 1.4cm 0.8cm},clip,scale=0.45]{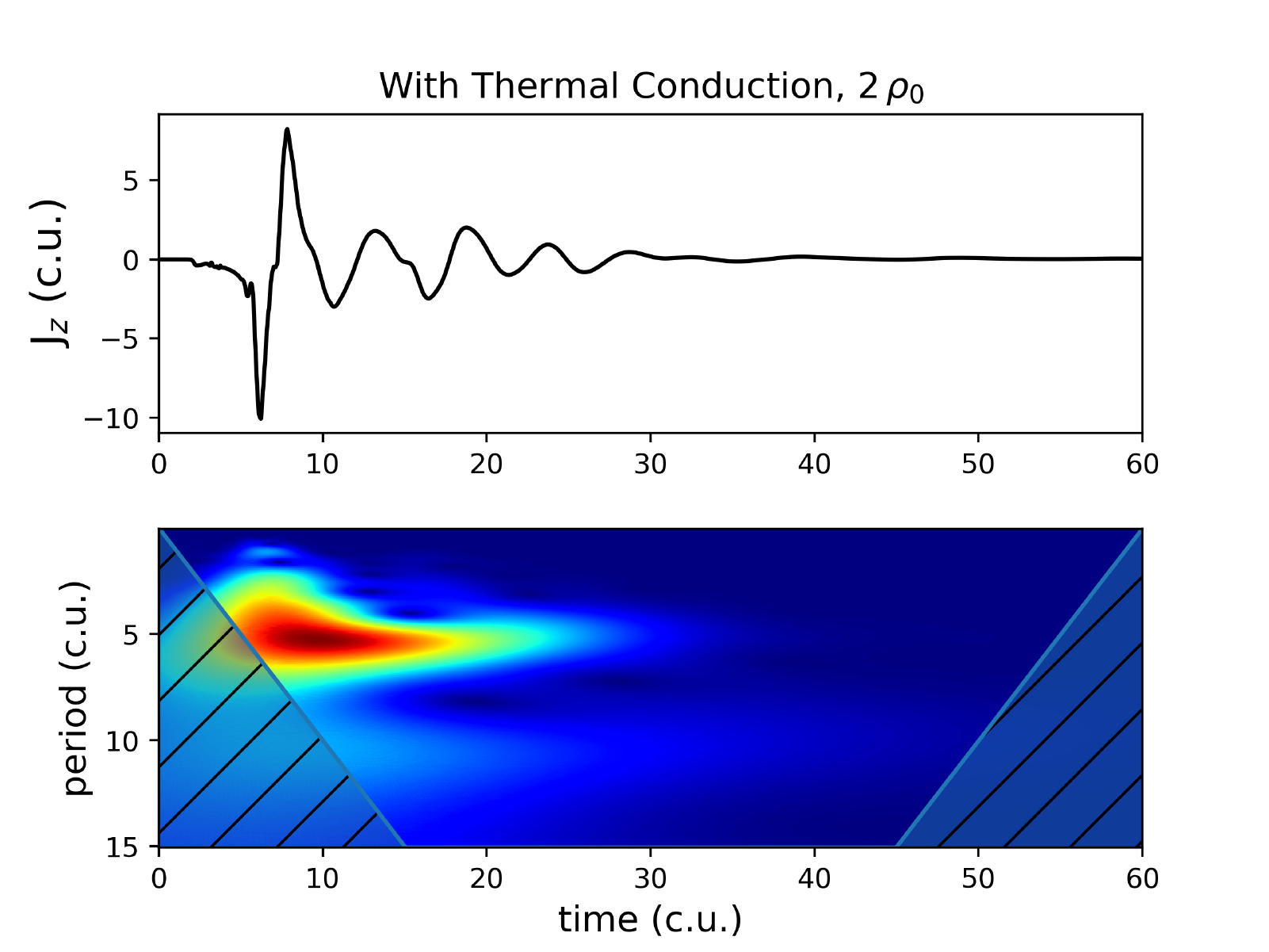}
    \includegraphics[trim={0.4cm 0.cm 1.4cm 0.8cm},clip,scale=0.45]{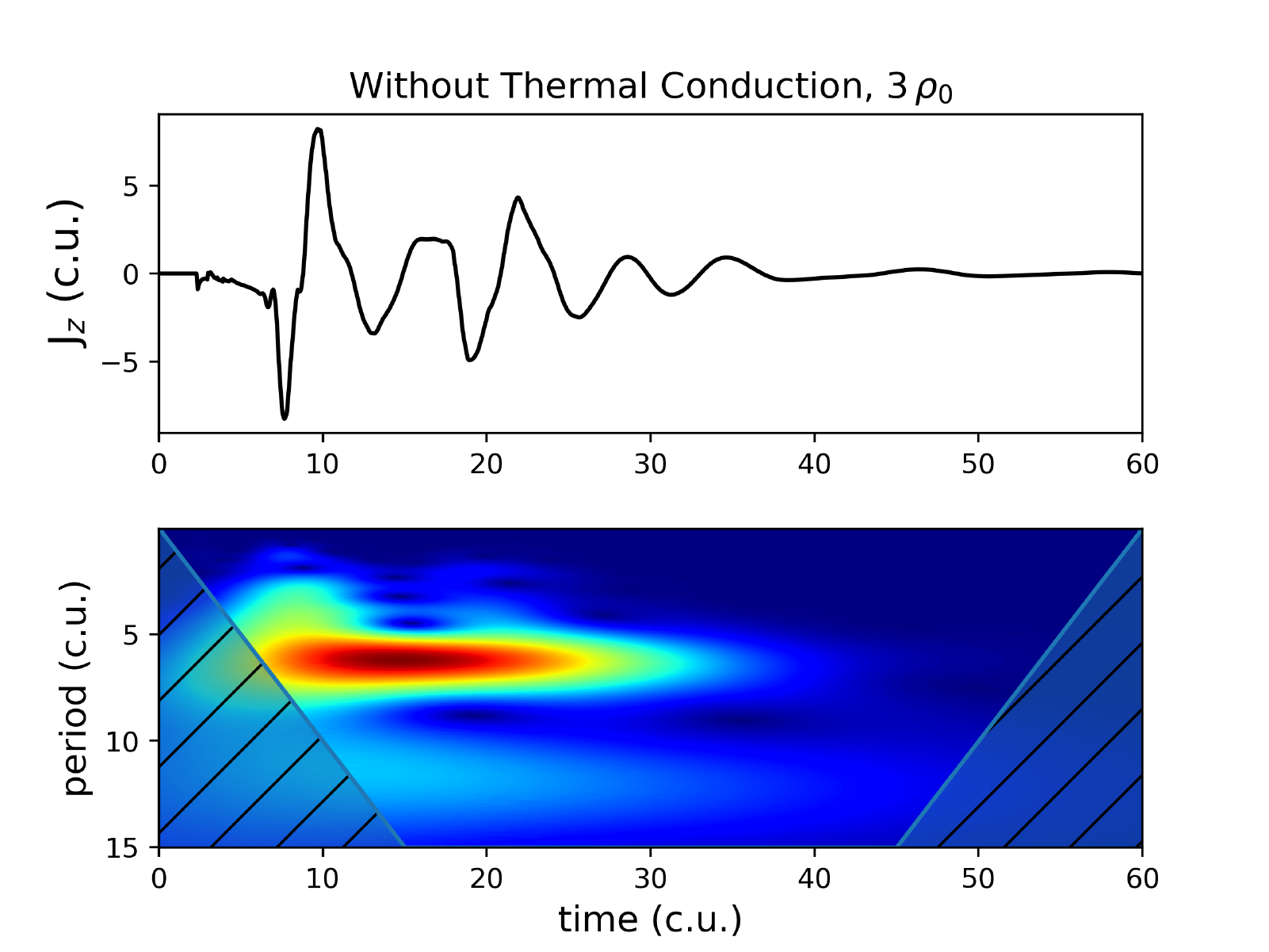}
    \includegraphics[trim={0.4cm 0.cm 1.4cm 0.8cm},clip,scale=0.45]{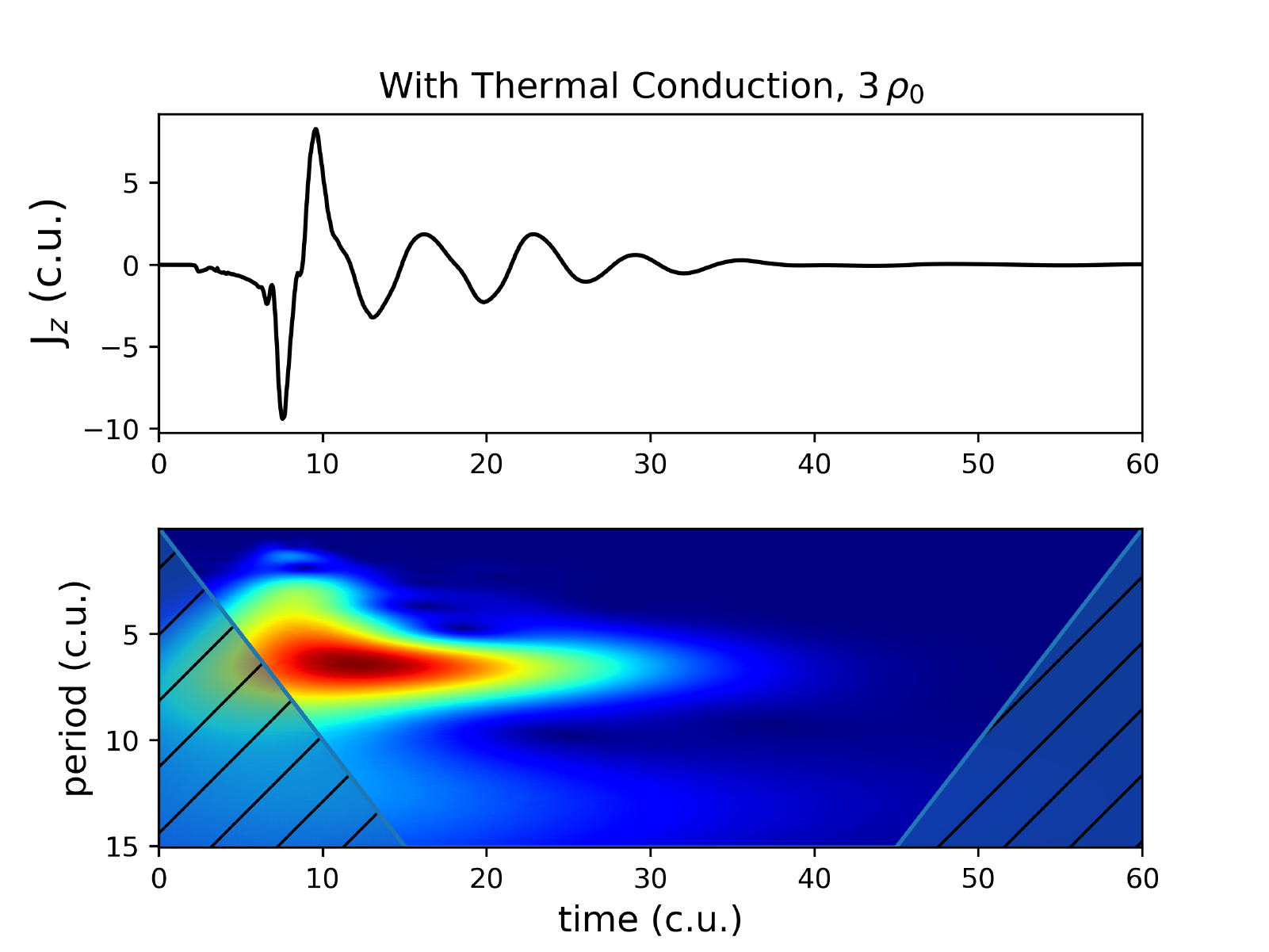}
	\includegraphics[trim={0.4cm 0.cm 1.4cm 0.8cm},clip,scale=0.45]{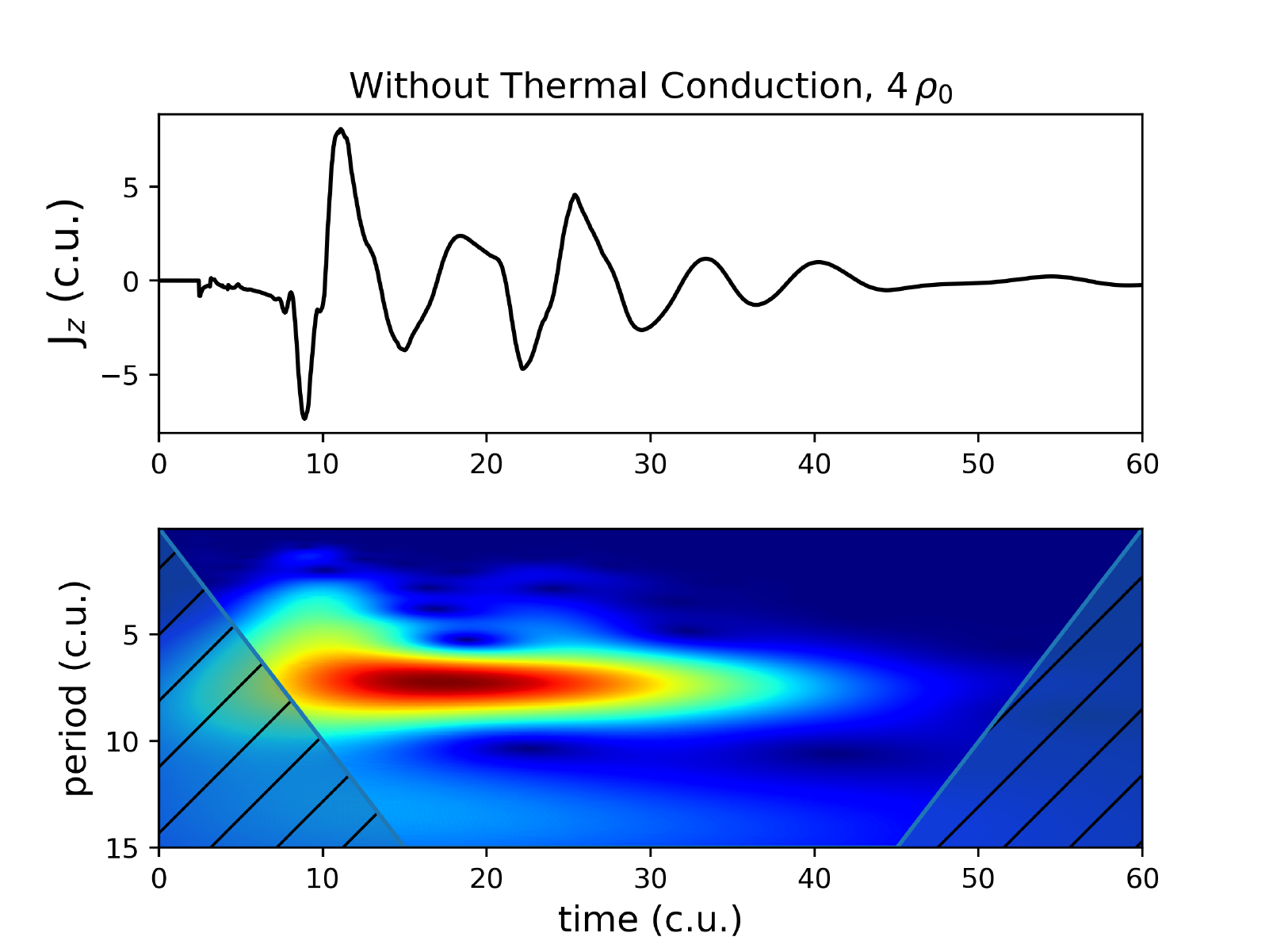}
    \includegraphics[trim={0.4cm 0.cm 1.4cm 0.8cm},clip,scale=0.45]{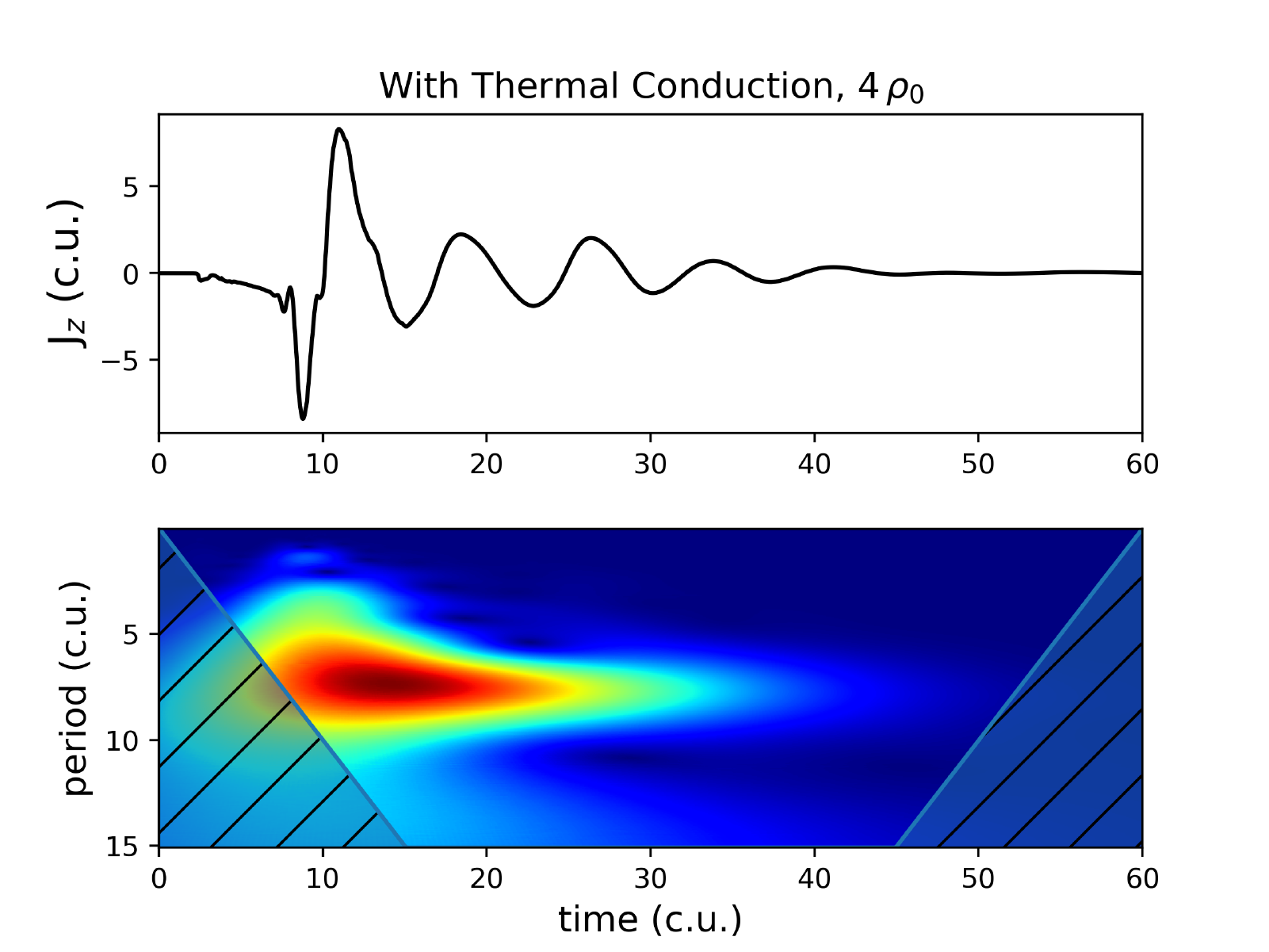}
    \caption{Time series of the $J_z$ current density at the null point for setups with different equilibrium density ($\rho=1,\,2,\,3$ and $4\,\rho_0$). The equilibrium magnetic field magnitude at a radius $r=1$ and temperature are $1\,B_0$ and $1$\,MK respectively. Again, cases without (left column) and with anisotropic thermal conduction (right column) are considered. All values are depicted in code units.}
    \label{fig:waveletR}
\end{figure*}

\begin{figure}[t]
    \centering
    \includegraphics[trim={0.cm 0.cm 0.cm 0.cm},clip,scale=0.45]{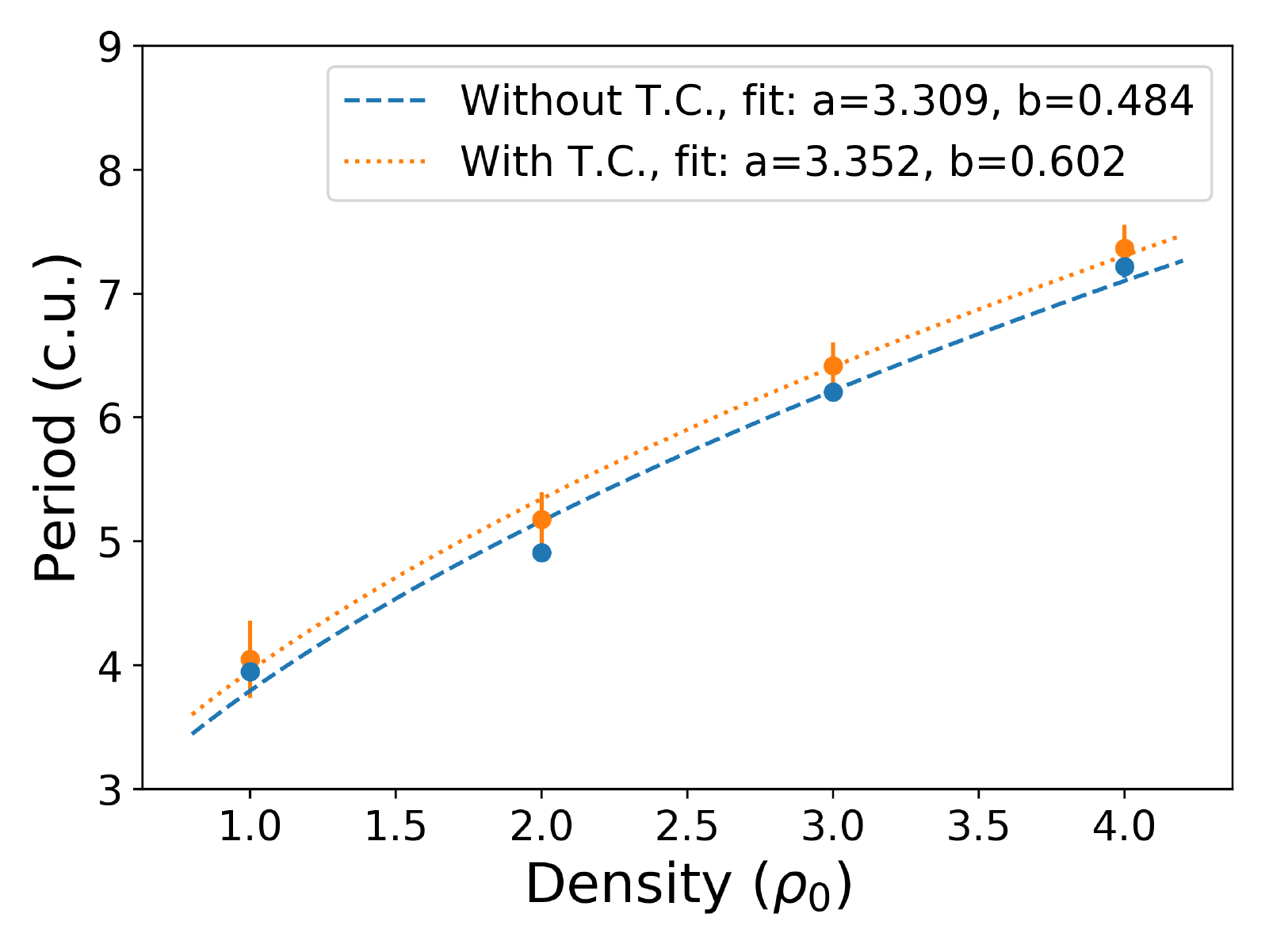}
    \caption{Same as Figure \ref{fig:PerMagn}, but here we depict the oscillation period with respect to the background density. Overplotted are the fits for both distributions of the function $G(\rho_0)=a\,(\rho_0)^{1/2}+b$.}
    \label{fig:PerDens}
\end{figure}

\begin{figure*}[t]
    \centering
    \resizebox{\hsize}{!}{
    \includegraphics[trim={0.4cm 0.cm 1.4cm 0.8cm},clip,scale=0.45]{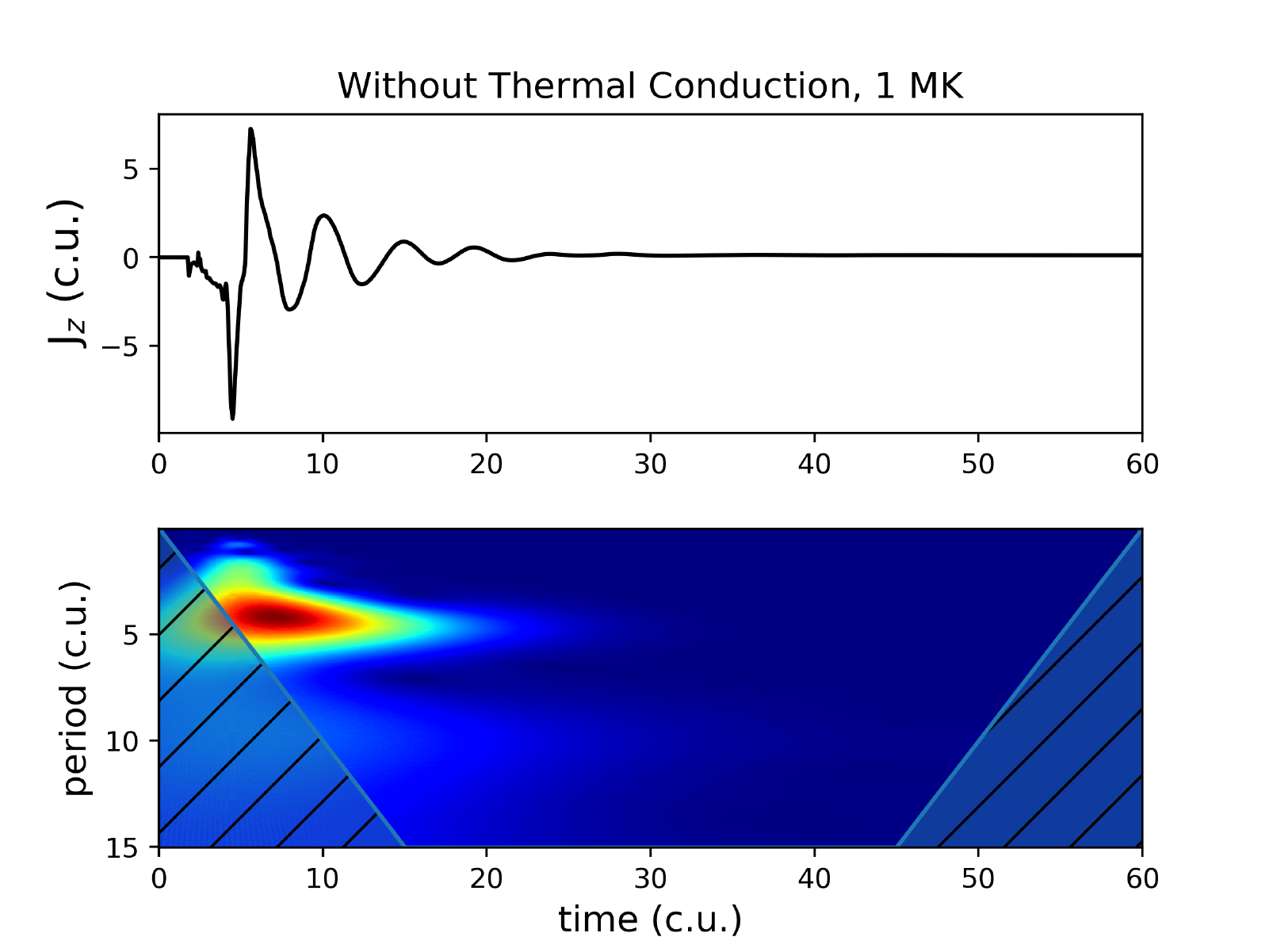}
    \includegraphics[trim={0.4cm 0.cm 1.4cm 0.8cm},clip,scale=0.45]{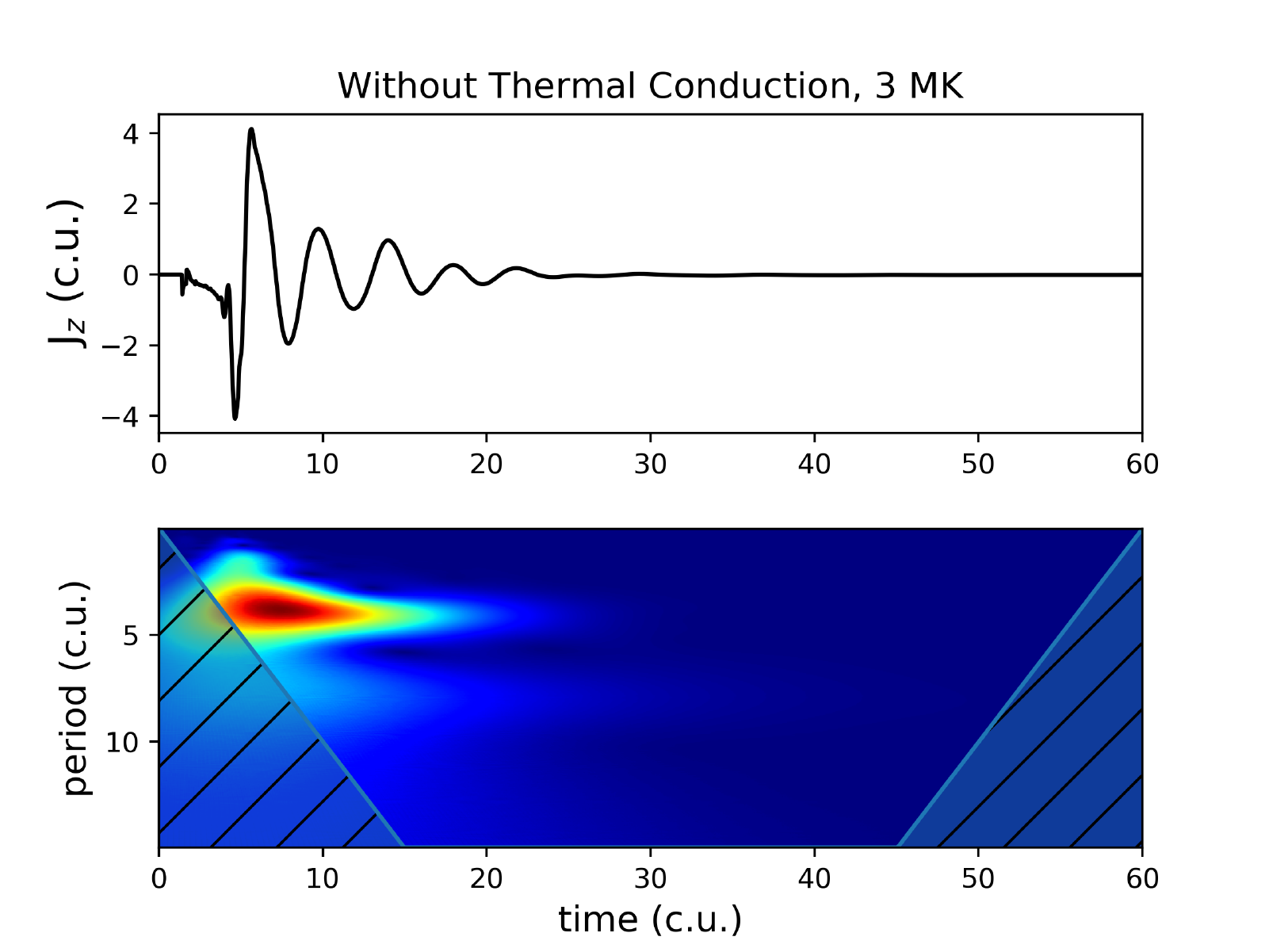}
    \includegraphics[trim={0.4cm 0.cm 1.4cm 0.8cm},clip,scale=0.45]{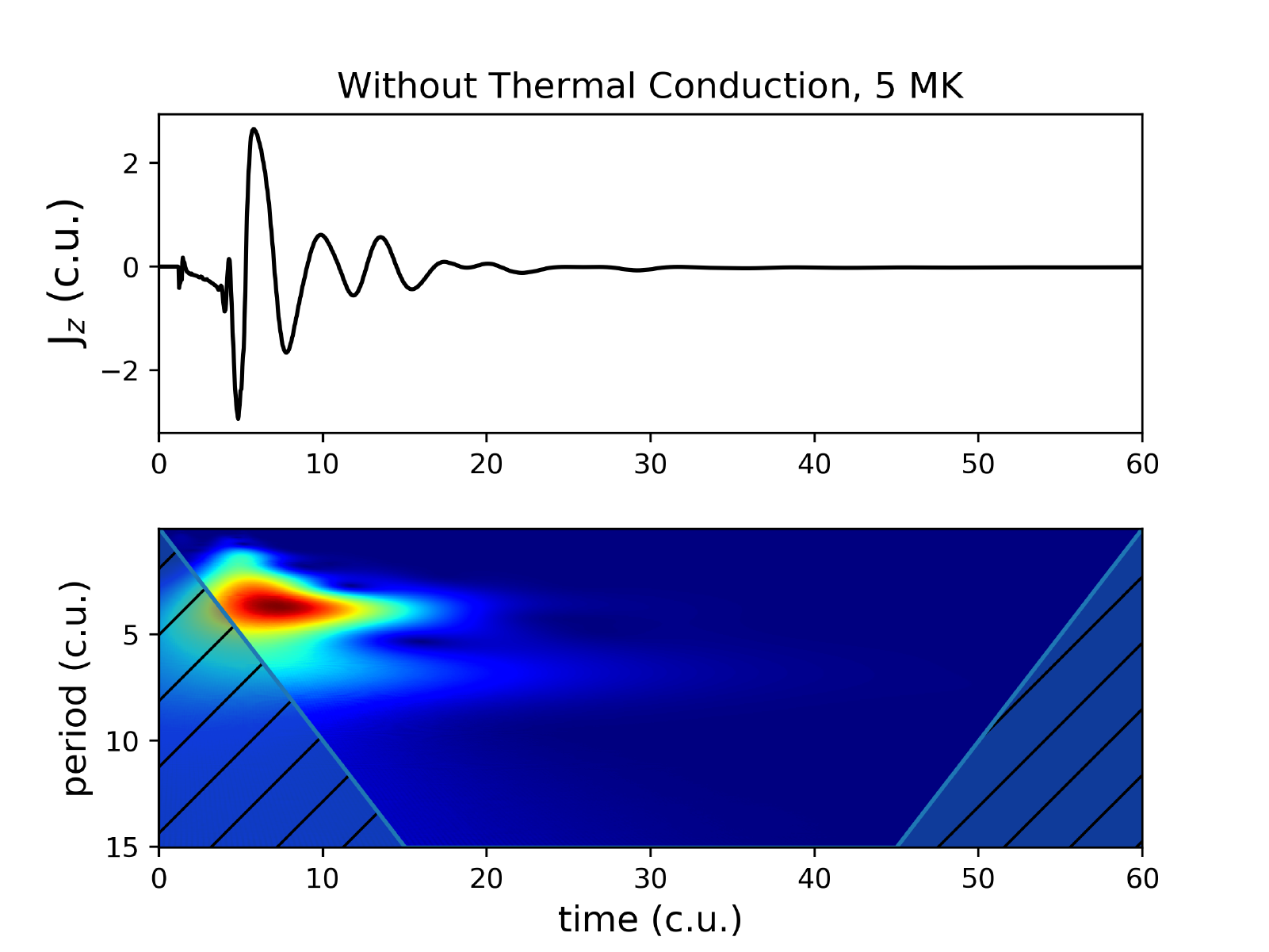}
    }
    \includegraphics[trim={0.4cm 0.cm 1.4cm 0.8cm},clip,scale=0.43]{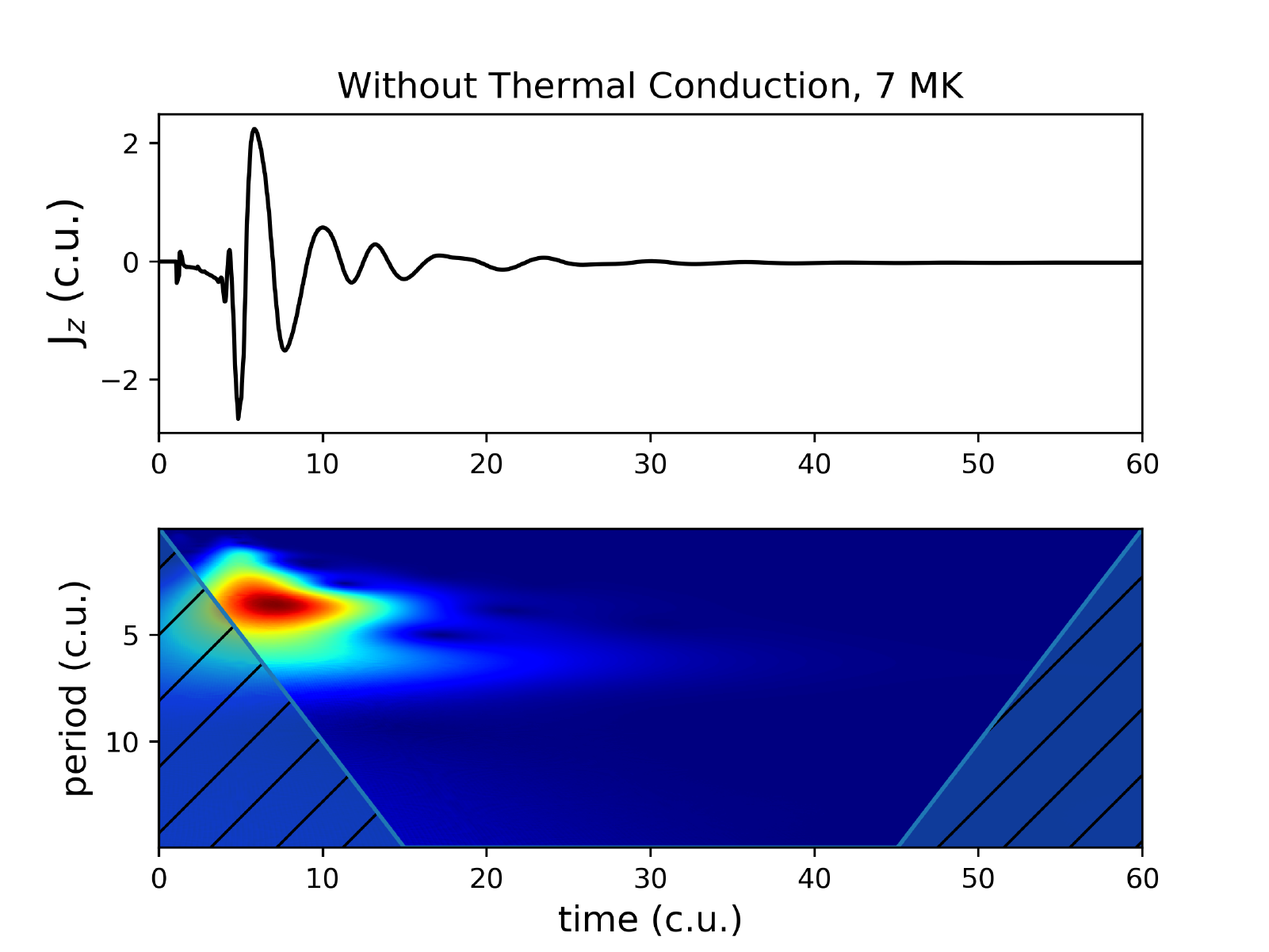}
	\includegraphics[trim={0.4cm 0.cm 1.4cm 0.8cm},clip,scale=0.43]{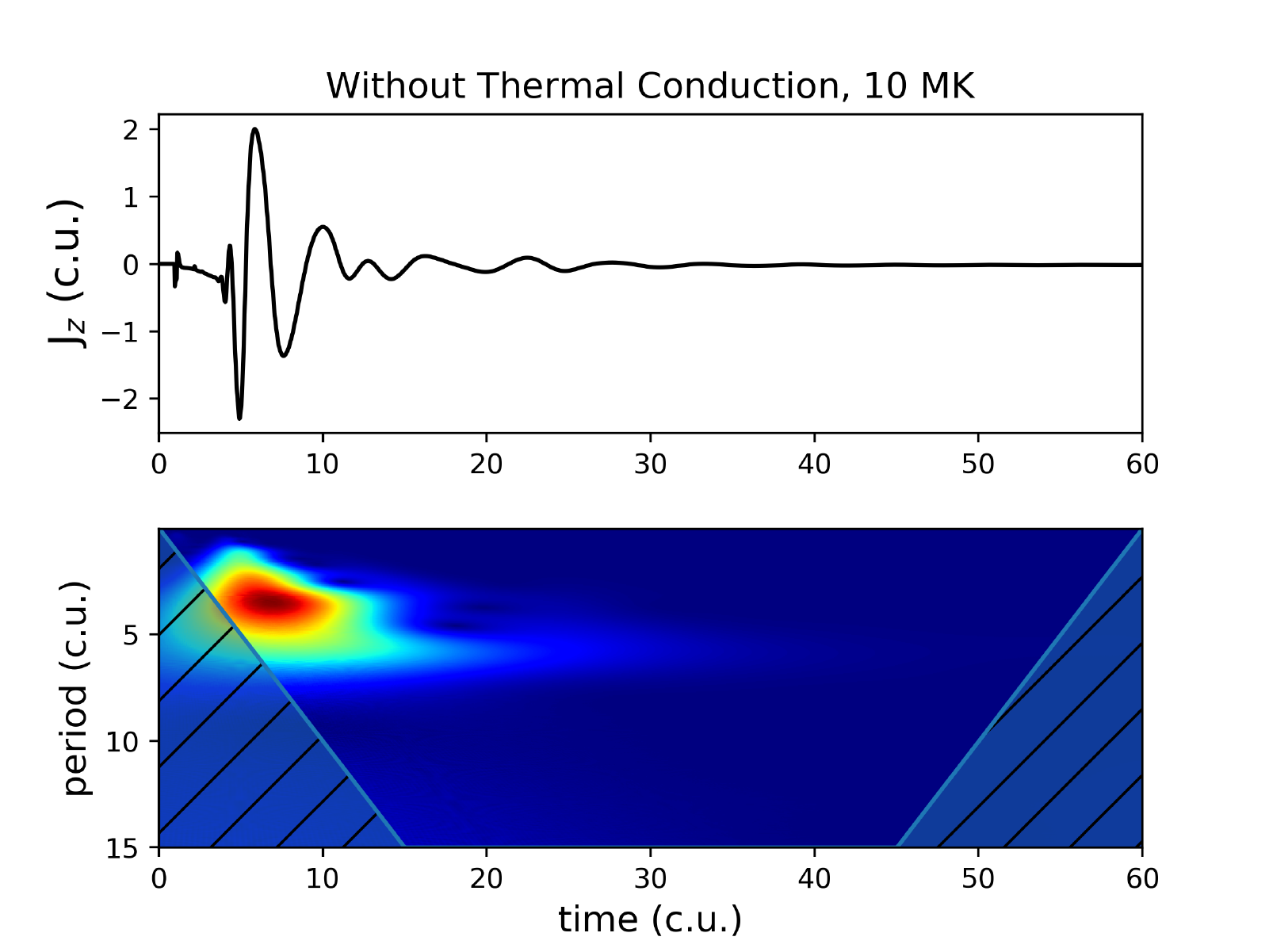}
    \caption{Time series of the $J_z$ current density at the null point for setups with different equilibrium temperature ($T=1,\,3,\,5,\,7$ and $10$\,MK). The equilibrium magnetic field magnitude at a radius $r=1$ and density are $1\,B_0$ and $1\,\rho_0$ respectively. Only setups without anisotropic thermal conduction are considered. All values are depicted in code units.}
    \label{fig:waveletΤ}
\end{figure*}

\begin{figure}[t]
    \centering
    \includegraphics[trim={0.cm 0.cm 0.cm 0.cm},clip,scale=0.45]{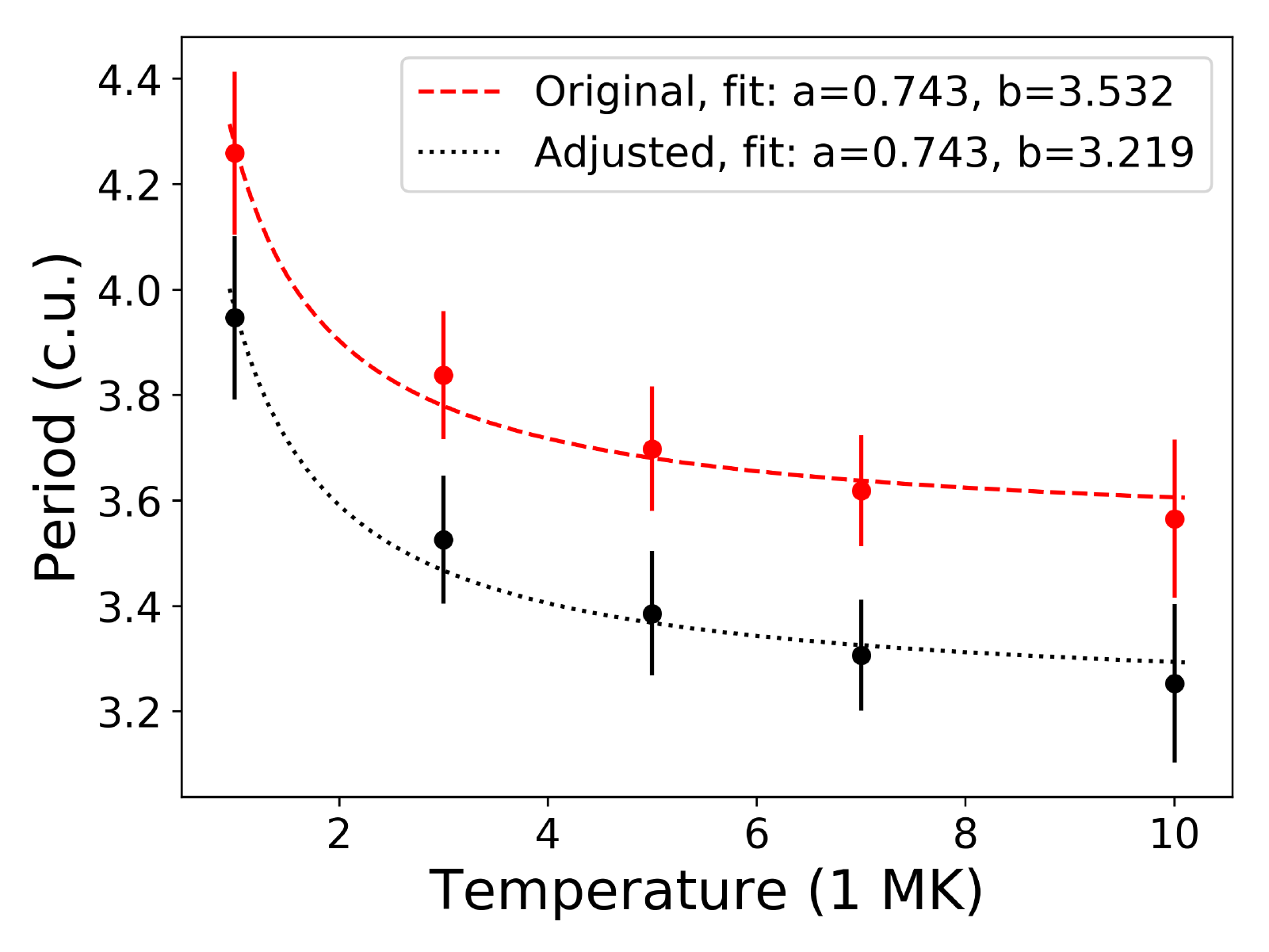}
    \includegraphics[trim={0.cm 0.cm 0.cm 0.cm},clip,scale=0.45]{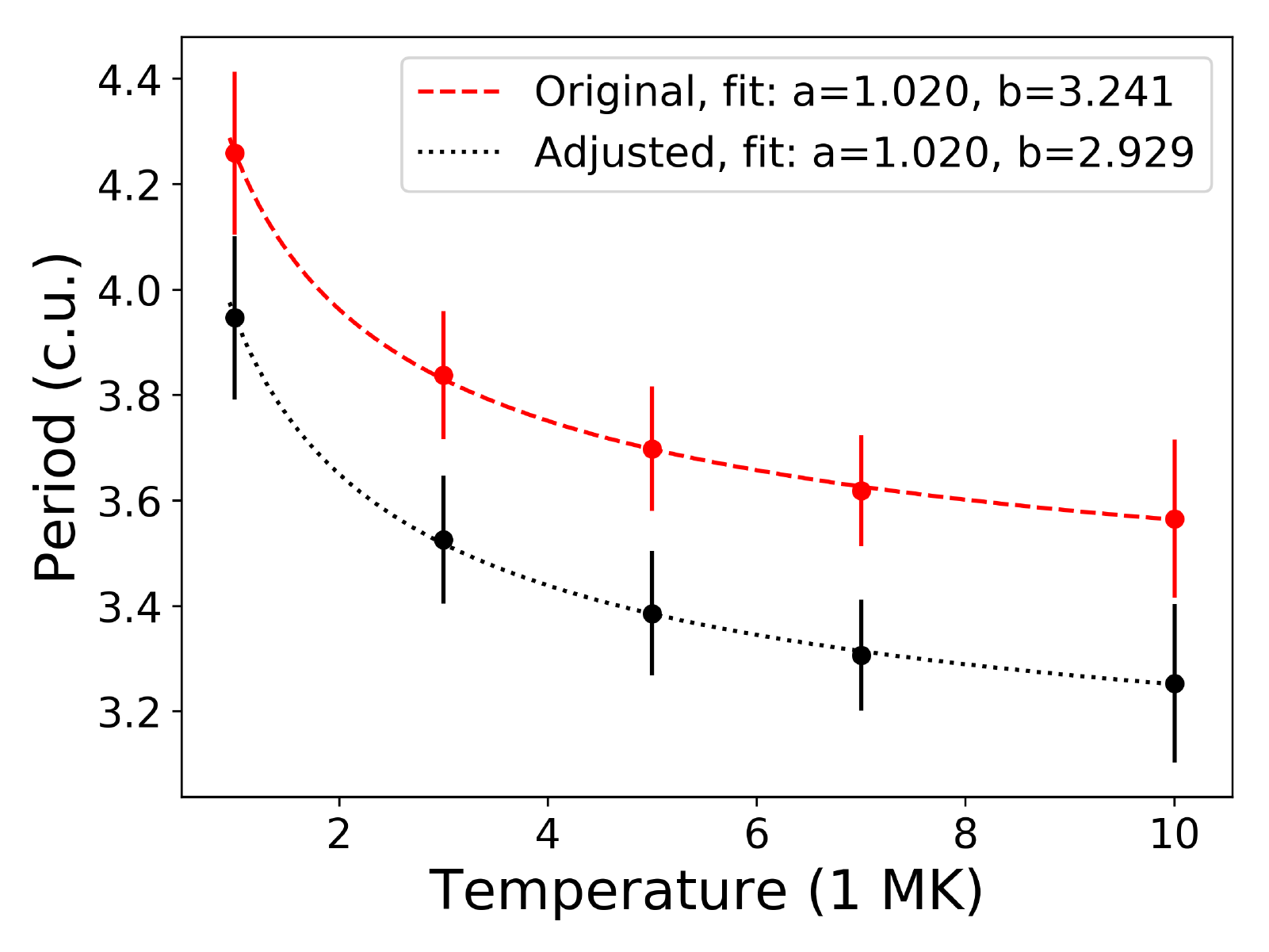}
    \caption{Graph depicting the distribution of the $J_z$ oscillation period with respect to the equilibrium temperature. In the left panel, we fit the function $F(T_0)= a\,(T_0)^{-1}+b$ to our original data points (red dashed line) and to the adjusted data points (black dotted line). In the right panel, we do the same but for the function $H(T_0)= a\,(T_0)^{-1/2}+b$. All setups are considered in the absence of thermal conduction. All values are depicted in code units, unless stated otherwise.}
    \label{fig:PerTemp}
\end{figure}

\begin{figure*}[t]
    \centering
    \includegraphics[trim={0.cm 0.cm 0.cm 0.cm},clip,scale=0.45]{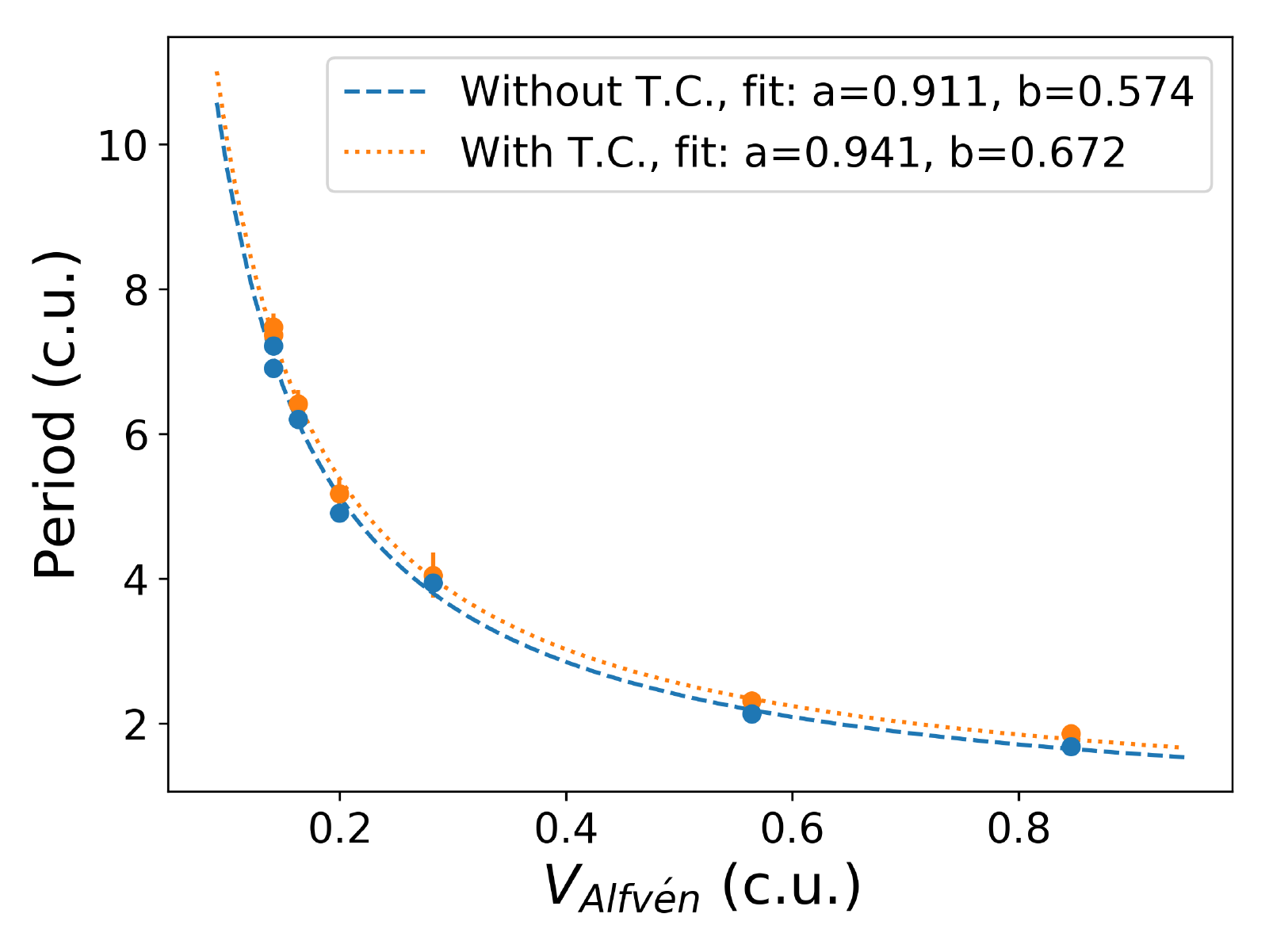}
    \includegraphics[trim={0.cm 0.cm 0.cm 0.cm},clip,scale=0.45]{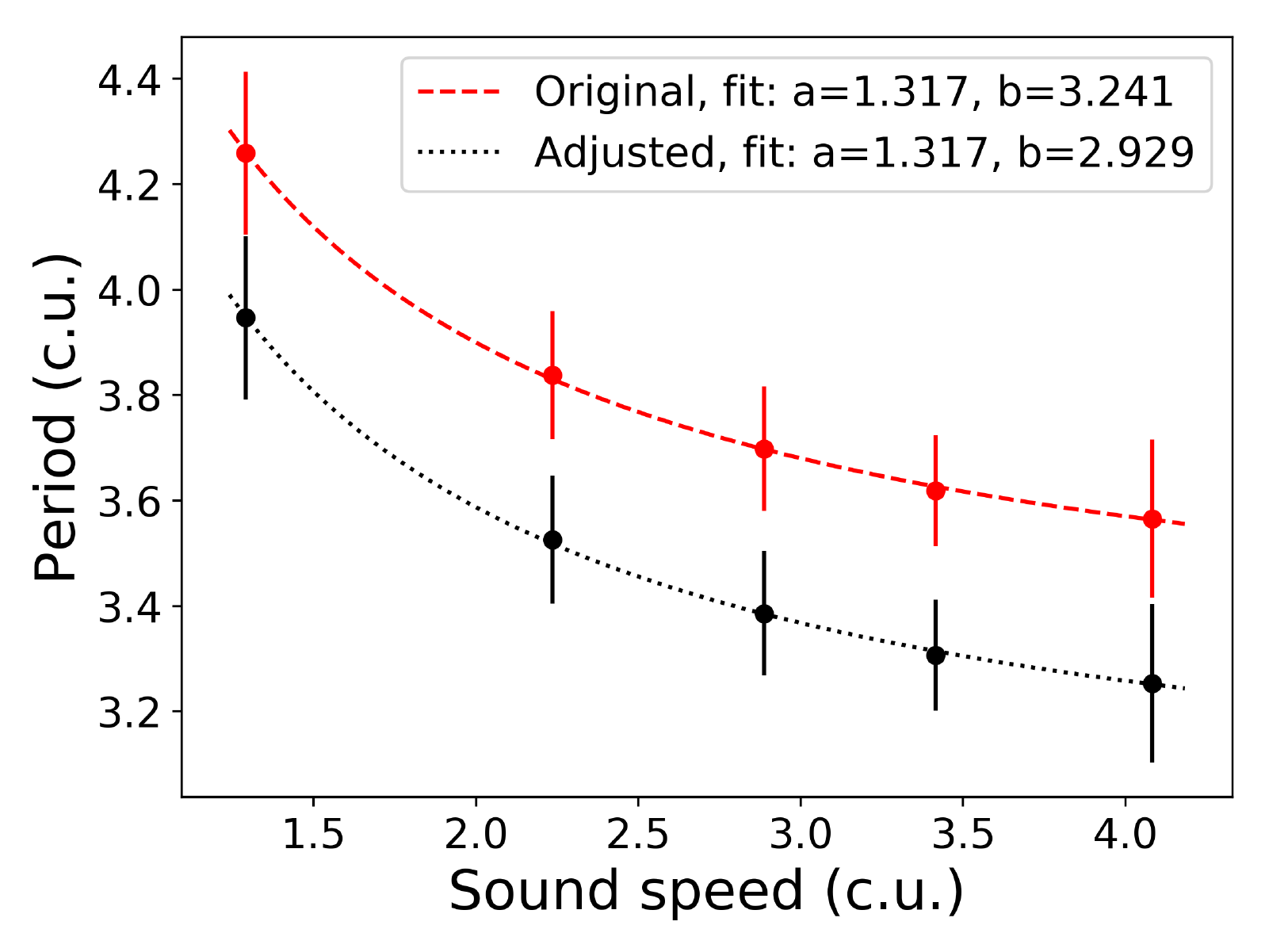}
    \caption{Left: Oscillation period versus the background Alfv\'{e}n speed at radius $r=1$, calculated for all the setups with the different background density and equilibrium magnetic field. The function $F(V_A)= a\,(V_A)^{-1}+b$ is fitted for both data sets. The color choice of Figure \ref{fig:PerMagn} is also followed here. Right: Same graph as in Figure \ref{fig:PerTemp}, only now the background sound speed is depicted instead of the background temperature. The function $F(V_S)=a\,(V_S)^{-1}+b$ is fitted for both data sets. All values are depicted in code units, unless stated otherwise.}
    \label{fig:PerAlfvSound}
\end{figure*}


\subsection{Numerical Scheme}\label{sec:numschm}

For the numerical studies below, we solve the 2D compressible resistive MHD equations in cartesian coordinates, in the absence of gravity \citep[see \S 2.1 in][]{Karampelas2022a}, using the PLUTO code \citep{mignonePLUTO2007, mignonePLUTO2012}. Like in our past studies \citep{Karampelas2022a,Karampelas2022b}, we employ the fifth-order monotonicity preserving scheme (MP5) for the spatial integration and the third-order Runge-Kutta method for the time integration. To satisfy the solenoidal constraint of the magnetic field ($\nabla\cdot\mathbf B=0 $), we use the Constrained Transport method  implemented in the code.

In these simulations, we also consider setups where we introduce anisotropic thermal conduction. The values for the parallel and perpendicular thermal conduction coefficients (in J\,s$^{-1}$\,K$^{-1}$\,m$^{-1}$), as calculated from the Spitzer conductivity \citep{Orlando2008ApJ}, are given below:
\begin{eqnarray}
\kappa_{\parallel} &=& 5.6 \times 10^{-12}\, T^{\frac{5}{2}},\label{eq:kpar}\\
\kappa_{\perp} &=& 3.3 \times 10^{-21}\, \frac{n_{H}^2}{\sqrt{T}B^2},\label{eq:kprp}
\end{eqnarray}
where $\kappa_{\parallel}$, $\kappa_{\perp}$ and the hydrogen number density $n_{H}$, temperature $T$ and magnetic field $B$ are all given in SI\footnote{In cgs, the thermal conduction coefficients (in erg\,s$^{-1}$\,K$^{-1}$\,cm$^{-1}$) are given as $\displaystyle \kappa_{\parallel} = 5.6 \times 10^{-7}\, T^{\frac{ 5}{ 2}}$ and $\displaystyle \kappa_{\perp} = 3.3 \times 10^{-16}\, \frac{ n_{H}^2}{ \sqrt{T}B^2}$.}. The effects of saturation are also taken into account for very large temperature gradients. The corresponding source term ($\nabla \cdot \mathbf{F}_c$) in the energy equation varying between the classical ($\mathbf{F}_{class}$) and saturated thermal conduction ($F_{sat}$):
\begin{equation}
\nabla \cdot \mathbf{F}_{c} = \nabla \cdot \left( \frac{F_{sat}}{F_{sat}+|\mathbf{F}_{class}|} \mathbf{F}_{class} \right)
\end{equation}
\begin{equation}
\mathbf{F}_{class} = \kappa_{\parallel} \mathbf{\hat{b}} \left(\mathbf{\hat{b}} \cdot \nabla T \right) + \kappa_{\perp} \left[ \nabla T - \mathbf{\hat{b}}\left( \mathbf{\hat{b}} \cdot \nabla T \right) \right]
\end{equation}
\begin{equation}
F_{sat} = 5\,\phi\,\rho\, V_{S,iso}^3,
\end{equation}
where $V_{S,iso} = \sqrt{p/\rho}$ is the isothermal sound speed, $\mathbf{\hat{b}}=\mathbf{B}/|\mathbf{B}|$ is the unit vector in the direction of magnetic field and $\phi$ is a free code parameter (with a default value of $0.3$). For zero magnetic field, $\mathbf{F}_{c}$ reduces to $\mathbf{F}_{c}= \kappa_{\parallel} \, \nabla T$.

During this analysis, we will be working in code units $U = U_{ph}\, U_0^{-1}$, with $U_{ph}$ being the physical quantities and $U_0$ the normalization units $U_0$. The constants $U_0$ are characteristic values, chosen for solar coronal plasma. We consider the unit length $L_0=1$\,Mm, unit density $\rho_0 = 10^{-12}$\,kg m$^{-3}$, and unit velocity $v_0=1.29\times10^{5}$\,m s$^{-1}$, equal to $V_S / \sqrt{\gamma}$ for coronal plasma at $1$\,MK. We also take the unit temperature $T_0=1\,$MK. The characteristic magnetic field and unit time are respectively $B_0 = \sqrt{\mu \rho_0 v_0^2} = 1.44$\,G and $t_0 = L_0 / v_0 = 7.78$\,s.

Since we want to solve the resistive MHD equations, we take the magnetic diffusivity in code units as $\eta = R_m^{-1} = 10^{-5}$, where $R_m = (v_0\,L_0)/\eta=10^5$ is the magnetic Reynolds number, assuming that the typical length and velocity scales of our system are respectively $L_0$ and $v_0$. Due to the finite size of our grid, our code also faces the effects of the `effective' numerical diffusivity, which prevents us from using $R_m$ values closer to those expected in the solar corona. Through a parameter study, this numerical diffusivity is estimated to be in the order of  $10^{-6}$ to $10^{-5}$.


\subsection{Initial Setup}\label{sec:initial}

This numerical study focuses on the perturbations of a 2D magnetic X-point. Similarly to \citet{Karampelas2022a}, the equilibrium magnetic field is defined in physical units as:
\begin{equation}\label{eq:magneticfield}
    \mathbf{B} = \frac{B_0}{L_0}\left(y,x,0\right).
\end{equation}
In Equation (\ref{eq:magneticfield}), $B_0$ is the characteristic magnetic field strength, and $L_0$ is the characteristic length scale of the magnetic field variations. A visual depiction of the magnetic field is shown in Figure~\ref{fig:profileB}, where the black solid and dashed lines depict the magnetic field lines in the regions with opposite polarities; the separatrices are in red. From Equation (\ref{eq:magneticfield}) we also see that the magnitude of the magnetic field is proportional to the radius $r=\sqrt{x^2+y^2}$.

We consider uniform equilibrium values for the density and temperature across the physical domain, obtaining a uniform initial sound speed
\begin{equation}
 V_S= \sqrt{\gamma}\,V_{S,iso} = \sqrt{\gamma p/\rho} = \sqrt{\gamma\,R\,T}
\end{equation}
where $\gamma=5/3$ is the ratio of the specific heats, and $R$ is the specific gas constant. This also results in a increasing Alfv\'{e}n speed 
\begin{equation}
V_A= \frac{B}{\sqrt{\mu_0 \, \rho}} = \frac{B_0}{L_0}\frac{r}{\sqrt{\mu_0\,\rho}},
\end{equation}
($\mu_0$ is the magnetic permeability of vacuum) as we move away from the X-point. Additionally, the choice of a uniform initial density distribution prevents the development of phase mixing in our setups (e.g. \citealt{heyvaerts1983}). Figure~\ref{fig:profileB} also depicts the equipartition layer, i.e. the layer where the ratio of $V_A$ over $V_S$ equals one. Given that the initial $V_S$ is constant in our setups, and $V_A$ is proportional to the magnitude of the magnetic field, and thus the radius, the equipartition layer will initially be a circle of radius, $r_{eq}$, where:
\begin{equation}\label{eq:equipart}
r_{eq} = \frac{L_0}{B_0} \sqrt{\frac{\gamma\,\rho\,R\, T}{\mu_0}}.
\end{equation}
From Equation (\ref{eq:equipart}) we see that the initial radius of the equipartition layer in our setups will be defined by the values initial uniform plasma temperature and density, and by the characteristic strength of the magnetic field.

In order to initiate oscillatory reconnection at the X-point, we use a circular fast magnetoacoustic pulse \citep[mentioned as Ring driver in][]{Karampelas2022b} to perturb the magnetic field from its equilibrium state. The horizontal components of the velocity pulse, as shown in Figure \ref{fig:profiledriver}, are calculated as follows:
\begin{eqnarray}
v_x = (v_{\parallel}B_x+v_{\perp}B_y)/(B_x^2+B_y^2),\\
v_y = (v_{\parallel}B_x-v_{\perp}B_y)/(B_x^2+B_y^2),
\end{eqnarray}
where $v_{\perp} = (\mathbf{v}\times \mathbf{B})\cdot\hat{\mathbf{z}}$ is a quantity related to the velocity component perpendicular to the magnetic field lines and $v_{\parallel}=\mathbf{v}\cdot\mathbf{B}$ is a quantity related to the velocity component parallel to the magnetic field lines. Following \citet{Karampelas2022a}, we consider a fast magnetoacoustic wave pulse (in code units) of the form:
\begin{eqnarray}
v_{\perp}(t=0) &=& \frac{1}{0.2 \sqrt{2\pi}}\exp\left( -0.5 \frac{(r-5)^2}{0.2}\right),\label{eq:ringvp} \\
v_{\parallel}(t=0)&=&0.\label{eq:ringvl}
\end{eqnarray}


\subsection{Domain and Boundary Conditions}\label{sec:boundary}

Our setup consists of a square domain with a structured uniform grid with a range $(x,y)\in \left[-10,10\right]$\, in code units, and resolution of $1801\times 1801$ grid points. We use reflective boundaries for the velocity components ($v_x,\,v_y$), so that no flows can cross the boundary and disrupt the initial equilibrium. To prevent the accumulation of heat at the boundaries, once thermal conduction is switched on, we fix pressure and density at the boundaries to their initial values. In order to keep the current density at the edges or our domain from getting artificial values due to boundary effects, we take zero-gradient boundary conditions for the magnetic field components (of the form $B_i - B_{i-1} = B_{i-1} - B_{i-2}$).

Following \citet{Karampelas2022b}, we take steps to minimize the amount of reflected waves from the boundaries returning to the null point. Our first step is to deal with the outward propagating velocity pulse that emerges from the splitting of the initial velocity annulus, as we can seen in Figure \ref{fig:pulevol} at t=$0.2$\,t$_0$. We do so, after the start of the each simulation, by turning the value of the velocity components to zero for a region with radius of $r\geqslant 7$.  

The second step is to create a numerical dissipation scheme away from the null point,
with the purpose of reducing the kinetic energy of the waves in that region. To that end, we divide each velocity component by a dissipation coefficient $n_d > 1$, for each iteration. The relation for the coefficient is given in code units:
\begin{equation}\label{eq:nd}
    n_{d}= 1.0005 +  0.0005\tanh(r - r_d), \qquad t>t_C,
\end{equation}
where $t_C$ is the time that we switch off the outward propagating pulse in the previous step and $r_d$ being the effective distance from the null point at which the scheme starts acting. The value for $r_d$ changes for each setup, to better accommodate the effects of the different Alfv\'{en} and sound speed profiles for each setup and to make the dissipation of the reflective waves more effective. 

Finally, for some of our setups we introduce explicit physical viscosity in the MHD equations, in addition to the previous numerical dissipation scheme \citep[see also][]{Karampelas2022b}, with coefficient in code units:
\begin{equation}\label{eq:nv}
    n_{visc}= 0.1 +  0.1\tanh(r - r_d), \qquad t>t_C.
\end{equation}


\section{Results} \label{sec:results}

The purpose of this study is to gain a better understanding into the nature of oscillatory reconnection in a hot coronal plasma and to explore its behaviour under different coronal conditions. To that end, we expand the results of \citet{Karampelas2022a} through a series of parameter studies. For each parameter study, we change either the characteristic strength of the magnetic field, the equilibrium density or the initial temperature. The studies for the different magnetic field and density have been performed both in the absence and presence of anisotropic thermal conduction, whereas the parameter study for the temperature has been performed only for setups without thermal conduction. An overview of the different cases can be found in Table \ref{tab:param}.


\begin{deluxetable}{ccccccc}
\tablewidth{0pt} 
\tablecaption{An overview of the physical parameters (in code units) for the different models in our simulations.\label{tab:param}}
\tablehead{ 
\colhead{Model} & \colhead{$B\,(B_0)$} & \colhead{$\rho\,(\rho_0)$} & \colhead{$T\,(T_0)$} & \colhead{$\kappa_{\perp},\kappa_{\parallel}$} & \colhead{$n_{visc}$} & \colhead{$r_d\,(L_0)$}}
\startdata 
$1$ & $ 0.5$ & $ 1.0$ & $1.0$ & $0$, $\neq 0$ & $\neq 0$ & $6$ \\ 
$2$ & $ 1.0$ & $ 1.0$ & $1.0$ & $0$, $\neq 0$ & $\neq 0$ & $5$ \\ 
$3$ & $ 2.0$ & $ 1.0$ & $1.0$ & $0$, $\neq 0$ & $\neq 0$ & $6$ \\ 
$4$ & $ 3.0$ & $ 1.0$ & $1.0$ & $0$, $\neq 0$ & $\neq 0$ & $6$ \\ 
$5$ & $ 1.0$ & $ 2.0$ & $1.0$ & $0$, $\neq 0$ & $\neq 0$ & $6$ \\ 
$6$ & $ 1.0$ & $ 3.0$ & $1.0$ & $0$, $\neq 0$ & $\neq 0$ & $6$ \\ 
$7$ & $ 1.0$ & $ 4.0$ & $1.0$ & $0$, $\neq 0$ & $\neq 0$ & $6$ \\ 
$8$ & $ 1.0$ & $ 1.0$ & $1.0$ & $0$ & $0$ & $5$ \\
$9$ & $ 1.0$ & $ 1.0$ & $3.0$ & $0$ & $0$ & $5$ \\
$10$ & $ 1.0$ & $ 1.0$ & $5.0$ & $0$ & $0$ & $6$ \\
$11$ & $ 1.0$ & $ 1.0$ & $7.0$ & $0$ & $0$ & $6$ \\
$12$ & $ 1.0$ & $ 1.0$ & $10.0$ & $0$ & $0$ & $6$ 
\enddata
\tablecomments{{Models 1 to 7 have been studied for both without ($\kappa_{\perp},\,\kappa_{\parallel} = 0$) and with ($\kappa_{\perp},\,\kappa_{\parallel} \neq 0$) anisotropic thermal conduction.}}
\end{deluxetable}

The initial velocity perturbation described by Equations (\ref{eq:ringvp}) and (\ref{eq:ringvl}) splits into two counter-propagating pulses of equal amplitude, with each travelling to opposing directions. While we deal with outward propagating pulse in the way that was described in the previous section, we focus on the evolution and effects of the pulse approaching the null point. The inward propagating pulse focuses at the X-point due to refraction, as shown in Figure \ref{fig:pulevol} for the default setup without thermal conduction ($B_0=1$, $\rho_0=1$ and $T=1$\,MK, see Model 2 from Table \ref{tab:param}). Mode conversion takes place as the fast magnetoacoustic wave pulse crosses the equipartition layer, from the region of low-$\beta$ to the region of high-$\beta$ plasma \citep[e.g.][]{McLaughlin2006b,Karampelas2022a}, deforming the layer in the process due to the formation of strong compression and rarefaction shocks in the $y$-direction and $x$-direction respectively \citep[see also][]{Gruszecki2011null}. 

Once the pulse reaches the null point, it perturbs it from its equilibrium, forcing it to perform a series of reconnection events, that are characterized by a periodic manifestation of horizontal and vertical current sheets (i.e. oscillatory reconnection). Like in the past studies, our main tool of studying oscillatory reconnection will be the tracking of the oscillating $J_z$ current density at the perturbed null point, as was first performed by \citet{McLaughlin2009}, and the calculation of its period for each different case.

\subsection{Magnetic Field Dependence} \label{sec:magfield}

Our first goal is to revisit the effects of the characteristic strength of the magnetic field ($B_0$) on oscillatory reconnection of an X-point in a hot coronal plasma. A first study has been performed in \citet{Karampelas2022a}, for an X-point in the presence of anisotropic thermal  conduction. Here we will repeat this analysis for the updated numerical dissipation scheme that was first introduced in \citet{Karampelas2022b}. The latter is more efficient in dealing with the reflections returning to the perturbed null point and thus leads to less contamination of the $J_z$ current density signal and a cleaner resulting spectrum. Unlike the previous parameter study on the magnetic field strength \cite[see][]{Karampelas2022a}, here we will expand the analysis for setups both in the presence and absence of anisotropic thermal conduction. In total we will consider four different values for the characteristic strength of the magnetic field ($0.5B_0,\, 1\,B_0,\, 2\,B_0$ and $3\,B_0$, where $B_0=1.44$\,Gauss). We note here that the magnitude of the magnetic field is proportional to the radius for the X-point, and that the characteristic value of the field is not the maximum value in our setups.  As we can see in Table \ref{tab:param} for models 1 to 4, in these four cases the initial density and temperature are $1\,\rho_0=10^{-12}$\,kg m$^{-3}$ and $1\, T_0=1$\,MK, and we will consider both the numerical dissipation scheme and a non-zero viscosity coefficient ($n_{visc}$) away from the null, in order to deal with the reflective waves. 

The produced time series for the $J_z$ current density of the different cases are shown gathered in Figure \ref{fig:waveletB}, were the results both and with and without thermal conduction are shown. Upon a visual inspection, we see that in all cases oscillatory reconnection has developed, as is hinted by the oscillatory $J_z$ signal at the null. The time series reveal for a stronger, and therefore stiffer magnetic field, the phenomenon of oscillatory reconnection lasts for progressively shorter times, before the oscillation is damped. This is in agreement with \citet{Karampelas2022a}, where it was shown that the decay rates for of these oscillations increase for stronger magnetic fields. On that note and to reduce the computational costs, the simulations for $2\,B_0$ and $3\,B_0$ are left running only up to $t=40\,t_0$, since the oscillation decays faster than the other cases. In the same study it was also shown that the period of the oscillation also decreases for stronger, stiffer magnetic fields. This can also be derived from Figure \ref{fig:waveletB}, once we focus on the calculated wavelet spectra for each case, shown here below their respective time series. As we see, there is a clear trend regarding the period of the oscillation, with the dominant period band being shifted towards smaller values for stronger fields. Finally, we see that for most of the cases studied here, the dissipation scheme used to deal with the reflected waves is working efficiently, allowing us to produce clear spectra, where there is one clearly defined band of periods. The only exception is for the case of $0.5\,B_0$ without anisotropic thermal conduction, where a strong secondary band of periods is observed. From \citet{Karampelas2022a} and  \citet{Karampelas2022b} it was shown that these secondary period bands are associated with the reflected waves returning to the null point. This means that for this particular case, with $0.5\,B_0$ characteristic magnetic field strength, our dissipation scheme was less effective than in the other cases. However, the main period band is still clearly defined and more prominent that the other one. 

In order to quantify this trend, we use the wavelet spectra to calculate the oscillation period for each case. We do so by first locating the coordinates (time $t_0$ and period $P_0$) of maximum power for each spectrum. We then consider a time interval $\Delta t=\left[ t_0-P_0,\,t_0+3\,P_0  \right]$ containing the periods that exhibit higher values of power, for which we calculate the average value for the period, and the standard deviation that will act as the error in the calculation. The calculated average values for the period of each oscillating signal are then placed in the graph of period versus the magnetic field strength, shown in Figure \ref{fig:PerMagn}. The calculated standard deviation for each value is added as error bars for each point, although for most cases theses error bars are barely visible. The data points on Figure \ref{fig:PerMagn} clearly hint towards an inverse proportionality relation between the oscillation period and the magnetic field strength. Because of this, we have fitted both sets of data points (with thermal conduction, in orange and without thermal conduction, in blue) with the function $F(B_0)=a\,(B_0)^{-1}+b$. Figure \ref{fig:PerMagn} also contains the values of the coefficients for both cases, which are $a=(3.159 \pm 0.096,\,3.398 \pm 0.046)$ and $b=(0.642 \pm 0.111,\,0.671 \pm 0.053)$ for the cases without and with thermal conduction respectively. We see that the addition of thermal conduction does not alter the trend in any significant way. We also see that the setups with thermal conduction generally give higher values of the period than those cases without thermal conduction, in agreement with our past studies \citep{Karampelas2022a,Karampelas2022b}.


\subsection{Density Dependence}\label{sec:density}

Our next goal is to study the response of oscillatory reconnection for different equilibrium density. We have considered again four different cases, where we take density values $\rho=1,\,2,\,3$ and $4\,\rho_0$ where $\rho_0=10^{-12}$\,kg m$^{-3}$. In all cases we have taken a characteristic strength of the magnetic field equal to $1\,B_0=1.44$\,Gauss, and temperature $1\, T_0=1$\,MK. All cases are studied both in the presence and absence of anisotropic thermal conduction. Just like before, we will again consider both the numerical dissipation scheme and a non-zero viscosity coefficient ($n_{visc}$) away from the null, to treat the reflections. The details of the different models (2, 5, 6 and 7) are shown on Table \ref{tab:param}.

The derived time series for the $J_z$ current density are shown in Figure \ref{fig:waveletR} alongside their respective wavelet spectra. Again, the spectra of the time series reveal a prominent period band for each case, associated with the oscillatory reconnection process, the secondary period bands from the reflected waves being of lower power. Again, upon a visual inspection we see that by increasing the value of the equilibrium density, the resulting period of the oscillation increases as well, again with thermal conduction leading to higher periods.

Following the same process as in the previous case, we derive the average values for the period, and the errors from the standard deviation for each case and we place them in the same oscillation period-density graph, in Figure \ref{fig:PerDens}. Again, we see a clear trend for each set of data points (with and without thermal conduction, shown in orange and blue respectively). For each set of data points, we fit the function $G(\rho_0)=a\,(\rho_0)^{1/2}+b$, that we believe is showing the best agreement with the observed trend of the values of the period. The values of the coefficients, as derived from the fit, are $a=(3.309 \pm 0.303,\,3.352 \pm 0.190)$ and $b=(0.484 \pm 0.479,\,0.602 \pm 0.300)$ without and with thermal conduction, respectively and are also shown Figure \ref{fig:PerDens}.


\subsection{Temperature Dependence}\label{sec:temperature}

The final parameter study that we want to perform revolves around the response of oscillatory reconnection to the initial background temperature. In the previous subsection, we took setups of different background densities, but we kept temperature the same for all cases, meaning that the sound speed was always the same for those cases. In this section, we will consider setups with different temperatures, and therefore different sound speeds as well. We have considered 5 different cases, corresponding to models 8 to 12 on Table \ref{tab:param}. In all models we have taken a characteristic strength of the magnetic field equal to $1\,B_0=1.44$\,Gauss, and an initial density of $1\,\rho_0=10^{-12}$\,kg m$^{-3}$, while the temperature takes values of $1,\,3,\,5,\,7$ and $10$\,MK. Unlike the two previous parameter studies, no anisotropic thermal conduction is considered in this one. This is due to the ever increasing computational costs once thermal conduction is considered, caused by a combination of the increasing temperatures and the Alfv\'{e}n speed profile for our given magnetic field geometry, making these simulations very costly to perform for the proper resolution. Additionally, for these five cases viscosity has been dropped from the artificial dissipation scheme dealing with the reflections. The viscous scheme was not working efficiently for the cases with higher temperatures and so we decided to drop it from the setups with the lower temperatures, for consistency. 

Figure \ref{fig:waveletΤ} shows the produced times series of the $J_z$ current density profiles at the perturbed null point, and the corresponding wavelet spectra for each profile. Unlike before, the changes in period here seem to be more subtle from one setup to the next. We see the gradual appearance of a secondary period band, which becomes increasingly stronger for higher temperatures, but without ever reaching the same power as the main period band. Given the more uneven $J_z$ signal for higher temperatures, it is safe to associate this secondary period band with the reflected waves from the boundaries, polluting the null point region and giving rise to more noisy signals. 

Following the same methodology as before for the magnetic field and the density, we again calculate the average values for the period of each case, and their respective errors through the standard deviation, and we plot then together in a graph, showing the relation between the period of oscillation and the background temperature. The results are shown in red in both panels of Figure \ref{fig:PerTemp}. In the left panel, just like before, we also fit the the function $F(T_0)=a\,(T_0)^{-1}+b$, with the coefficients taking the values $a=0.743\pm 0.060$ and $b=3.532\pm 0.029$. Although, the fitted function passes through all the data points if we consider their error bars, we have also decided to fit the function $H(T_0)=a\,(T_0)^{-1/2}+b$ in our data (right panel), with coefficients $a=1.02\pm 0.013$ and $b=3.241\pm 0.007$. As we can see by comparing both panels of Figure \ref{fig:PerTemp}, the function $H(T_0)$ provides a better fit on the given data, with the coefficients presenting smaller errors by comparison to those for $F(T_0)$. Therefore, from now on we will be using the $H(T_0)=a\,(T_0)^{-1/2}+b$ function to describe the dependency of the period to the background temperature. Also, it becomes obvious that for our range of chosen temperature that matches the coronal conditions, the variations of the period are considerably smaller than the other case that we have examined. 

Finally, we need to address the effects of the different dissipation scheme used in this parameter study. As mentioned earlier, for the cases considered in this subsection we took the numerical dissipation scheme described by the coefficient of Equation (\ref{eq:nd}), without the supplementary viscous scheme described by the coefficient of Equation (\ref{eq:nv}). In other words, for models 8 to 12 of Table \ref{tab:param}, we took $n_{visc}=0$. When comparing the resulting periods for model 2 ($P=3.947 \pm 0.022$), used in the previous two subsections and from its equivalent model 8 used here ($P=4.259 \pm 0.155$), we see that the two produce slightly different results. It is not certain how removing the artificial viscous scheme leads to this small difference in period, of the order of $\Delta P \approx 0.312\, t_0 = 2.43$\,s. It is very likely that we are dealing with some code-specific numerical effects at this point, which would be hard to properly treat within the context of this study. However, the very small value of this difference makes us confident to compare our current results with those of the previous sections. To that end, we have subtracted the difference $0.312$ from the periods for all of our data points shown in both panels of Figure \ref{fig:PerTemp}. We do this, because none of the cases studied in this subsection had the viscous dissipation scheme switched on and thus we have been consistent among these five different setups. The resulting adjusted points (in black) follow the same trend as before for each panel, with the fitted function $F(T_0)$ having coefficients $a=0.743\pm 0.060$ (same as before) and $b=(3.532-0.312)\pm 0.029=3.219\pm 0.029$, and the fitted function $H(T_0)$ having coefficients $a=1.02\pm 0.013$ (same as before) and $b=2.929\pm 0.007$. We repeat that from now on we will be using the $H(T_0)=a\,(T_0)^{-1/2}+b$ function to describe the dependency of the period to the background temperature.


\section{Discussion and Conclusions} \label{sec:discussions}
\subsection{Parameter Studies}
In this paper we once again revisit the mechanism of oscillatory reconnection of a 2D X-point in hot coronal plasma, further exploring its response to different coronal conditions. This comes as a need, due to the large number of observations that can be attributed to the process of oscillatory reconnection. The first step was taken in \citet{Karampelas2022a} where the periodicity and the decay rate of the mechanism was studied in the presence of anisotropic thermal conduction in coronal conditions, expanding past studies that focused in cold plasma. In that same study, a clear connection was revealed between the magnetic field strength and the period of the oscillation. The second step was taken in \citet{Karampelas2022b}, where it was found that the period of oscillatory reconnection of a magnetic X-point perturbed by an external pulse is independent of the amplitude and type of the perturbing pulse. These two studies had already hinted towards the possibility of using oscillatory reconnection as a tool for coronal seismology. To that end, in the current study we have expanded upon the results of \citet{Karampelas2022a}, by considering cases with different magnetic field, density and background temperature. 

Our focus on the three quantities mentioned previously is based on the assumption that the properties of oscillatory reconnection, like its period, in the absence of dedicated external driving, should be related to the conditions of the background plasma in the vicinity of the null point. This is due to oscillatory reconnection being a fundamental process, related to the relaxation of a perturbed magnetic null point (here, X-point). Therefore, we do not expect the large scale magnetic field topology to affect our results, as the field geometry used here is characteristic of the field geometry in the immediate neighbourhood of an X-point. This is analogous to null-points acting as resonant cavities \citep[see][]{Santamaria2018} where the wave-null point interaction properties are determined by the background plasma properties near the null point. Here we need to note that having external driving would lead to a dependancy of the oscillation period to the period of the driving \citep[][]{Heggland2009ApJ}. However, our focus in this study is on the non-driven, relaxation based oscillatory reconnection.

Using the PLUTO code, we have solved the compressible and resistive 2D MHD equations, for a series of parameter studies. The first one included four different setups, each for a different value of the characteristic strength of the magnetic field ($0.5B_0,\, 1\,B_0,\, 2\,B_0$ and $3\,B_0$, where $B_0=1.44$\,Gauss), studied both with and without anisotropic thermal conduction. We note here that the characteristic value of the field is not the same as its maximum value in each setups, rather the field magnitude is proportional to the distance from the X-point. The results revealed an inverse proportionality between the period and the strength of the magnetic field, as shown in Figure \ref{fig:PerMagn}. In the second one we considered again four different cases, where we took the density to be equal to $\rho=1,\,2,\,3$ and $4\,\rho_0$ where $\rho_0=10^{-12}$\,kg m$^{-3}$, again studied both with and without anisotropic thermal conduction. The results, as shown in Figure \ref{fig:PerDens} reveal a square root relation between the period and the equilibrium density. The third and final parameter study involved five setups with different values of background temperature ($1,\,3,\,5,\,7$ and $10$\,MK), all studied in the absence of anisotropic thermal conduction. This last parameter study, shown in Figure \ref{fig:PerTemp}, revealed an inverse proportional relation, this time between the period and the square root of the background temperature.

As expected from our previous studies, the cases with anisotropic thermal conduction practically follow the same trend as their respective ones without thermal conduction, their only difference being that the former exhibit slightly higher values of period. The only exception are for those setups in the temperature parameter study, where thermal conduction has not been considered, due to increased computational costs for our given resolution. This is caused by a combination of the increasing temperatures and the given magnetic field geometry. 

Also, as was explained in the previous section, the derived values for the period from the temperature parameter study are shifted with respect to the rest, due to the slightly different artificial dissipation schemed, without any supplementary viscosity-based scheme that was used throughout it. Comparing the resulting periods of two equivalent setups, each with a version of the dissipation scheme, we get a difference of $\Delta P \approx 0.312\, t_0 = 2.43$\,s, which is of the order of $\sim 8\%$ from the value of $P\sim 30$\,s that we get from the other two parameter studies. Since we have used the same dissipation scheme when studying the response of oscillatory reconnection to temperature, we subtract $\Delta P$ from all of these results, as shown by the black line of the adjusted fit in both panels of Figure \ref{fig:PerTemp}. This allows for a better comparison with the other two parameter studies presented here. A synopsis of the fitted functions and the values of their coefficients can be found in Table \ref{tab:functions}.

\begin{deluxetable}{cccc}
\tablewidth{0pt} 
\tablecaption{Summary of fitted functions and the values of their coefficients for the parameter studies, as well as for the Alfv\'{e}n and sound speeds.\label{tab:functions}}
\tablehead{ 
\colhead{Study} & \colhead{Fit} & \colhead{$a$, $b$} & \colhead{$a$, $b$}\\ \colhead{} & \colhead{} & \colhead{(without T.C.)} & \colhead{(with T.C.)}}
\startdata 
Magnetic field   & $F(B_0)$    & $ 3.159$, $ 0.642$ & $ 3.398$, $ 0.671$  \\ 
Density          & $G(\rho_0)$ & $ 3.309$, $ 0.484$ & $ 3.352$, $ 0.602$  \\ 
Temperature      & $H(T_0)$    & $1.020$, $3.241$ & $\mathit{1.020}$, $\mathit{2.929}$  \\ 
Alfv\'{e}n speed & $F(V_A)$    & $ 0.911$, $ 0.574$ & $ 0.941$, $ 0.672$  \\
Sound speed      & $F(V_S)$    & $ 1.317$, $ 3.241$ & $\mathit{1.317}$, $ \mathit{2.929}$  \\ 
%
%
\enddata
\tablecomments{The three types of fitted functions are $F(x)=a\,(x)^{-1}+b$, $G(x)=a\,(x)^{1/2}+b$ and $H(x)=a\,(x)^{-1/2}+b$. For the temperature and sound speed, both pairs of coefficients are without thermal conduction, with the second pair being for the adjusted data sets. This is indicated in italics.}
\end{deluxetable}

\subsection{Period vs Alfv\'{e}n and Sound Speed}
Continuing on the trend set by our analysis so far, we want to study the evolution of the period of oscillatory reconnection in terms of the Alfv\'{e}n and sound speed profiles. The initial Alfv\'{e}n speed profile is dependent both on the initial equilibrium density and the characteristic magnetic field strength. We then take the results from the first 7 models of Table \ref{tab:param}, for the different values of density and characteristic magnetic field strength, calculate the characteristic Alfv\'{e}n speed and plot them with respect to the period. This graph is featured on the left panel of Figure \ref{fig:PerAlfvSound}. We again fit the function $F(V_A)=a\,(V_A)^{-1}+b$, the coefficients of which take values $a=(0.911\pm0.027,\,0.941\pm0.020)$ and $b=(0.574\pm0.136,\,0.672\pm0.103)$ for the datasets without and with thermal conduction, shown in dashed blue and dotted orange lines. This fit for the Alfv\'en speed is in agreement with the previous fits for the magnetic field and the density, since the Alfv\'en speed is proportional to the magnetic field and the inversely proportional to the square root of density. The right panel of the same figure, shows the results for the sound speed, which are derived from those of the temperature parameter study without thermal conduction. On that panel we show both the original values of the period (points in red) and the adjusted ones (points in black) for which we subtracted the difference $\Delta P \approx 0.312\, t_0 = 2.43$\,s as was mentioned in the section for the temperature parameter study. Finally, we have fitted the function $F(V_S)=a\,(V_S)^{-1}+b$, for the original (red dashed line) and the adjusted data (black dotted line), the coefficients of which take values $a=(1.317\pm0.016,\,1.317\pm0.016)$ $b=(3.241\pm0.007,\,2.929\pm0.007)$ for the original and adjusted data respectively. This fit agrees with the one of the $H(T_0)=a\,(T_0)^{-1/2}+b$ function for the background temperature, presented in Section \ref{sec:temperature}, since the sound speed is proportional to the square root of the temperature. This further justifies the use of function $H(T_0)$ to describe the relation between the period of oscillatory reconnection and the plasma temperature.  

\begin{deluxetable}{ccccccc}
\tablewidth{0pt} 
\tablecaption{Examples of calculating the period of oscillatory reconnection through Equation (\ref{eq:PnoTCb}).\label{tab:diagnostic}}
\tablehead{ 
\colhead{$B_{ph}$\,(G)} & \colhead{$\rho_{ph}$\,(kg m$^{-3}$)} & \colhead{$T_{ph}$\,(MK)} & \colhead{$P_{ph}$\,(s)}}
\startdata 
%
%
$10$ & $ 2.0\times 10^{-12}$ & $ 5.0$ & $15.9$  \\ 
$20$ & $ 2.0\times 10^{-12}$ & $ 5.0$ & $14.2$  \\ 
$30$ & $ 2.0\times 10^{-12}$ & $ 5.0$ & $13.6$  \\ 
$20$ & $ 2.0\times 10^{-12}$ & $ 3.0$ & $15.2$  \\
$20$ & $ 2.0\times 10^{-12}$ & $ 10.0$ & $13.1$  \\ 
$20$ & $ 3.0\times 10^{-12}$ & $ 5.0$ & $22.4$  \\ 
$20$ & $ 5.0\times 10^{-12}$ & $ 5.0$ & $35.3$  \\
$20$ & $ 30.0\times 10^{-12}$ & $ 5.0$ & $118.8$  \\
\enddata
\tablecomments{The subscript `\textit{ph}' refers to the physical quantities $U_{ph} = U\,U_0$, with $U$ the quantities in code units and $U_0$ the normalization unit (see also \S\ref{sec:numschm}).} 
\end{deluxetable}

\subsection{Empirical Formula}\label{sec:discussions3}
As a last step, we want to merge the derived relations from each one of the three parameter studies into one. We do this because one of the main goals of this present study was to start developing its capabilities as a plasma diagnostic tool. To that end, we are aided by the results of \citet{Karampelas2022b}, that allow us to ignore the strength of the perturbing pulse from this relation. Taking the cases without thermal conduction, we can merge the derived relations of each previous fit into the following  formula for our four key parameters:
\begin{equation}\label{eq:PnoTC}
\frac{P_{ph}}{t_0} = \frac{3.159\, B_0}{B_{ph}} + 3.309 \sqrt{\frac{\rho_{ph}}{\rho_0}} + 1.02\sqrt{\frac{T_0}{T_{ph}}} - 3.541\pm 0.434
\end{equation} 
where the penultimate right-hand term comes from solving the above equation for a known value of the period $P_{ph}$ (in s). For this, we considered the resulting period for model 2 ($P_{ph}=(3.947 \pm 0.022)\,t_0$). We also include the maximum error in the last right-hand term that is derived from the different combinations of errors in the values of $P_{ph}$ and the coefficients of the fits. Here, the subscript `\textit{ph}' refers to the physical quantities $U_{ph}$, as defined in \S\ref{sec:numschm}. For the quantities in physical units we have $U_{ph} = U\,U_0$, with $U$ the quantities in code units and $U_0$ the normalization units. We used the adjusted results for the temperature parameter study, as explained earlier, while all the coefficients are given in code units. Using the normalization units defined in \S\ref{sec:numschm}, we can rewrite the above formula in SI units, except for the magnetic field which is given in Gauss:
\begin{equation}\label{eq:PnoTCb}
P_{ph} = \frac{35.39}{B_{ph}} + 25.74 \, \times 10^{6} \sqrt{\rho_{ph}} + \frac{7.94}{\sqrt{T_{ph}}} - 27.55\pm 3.38
\end{equation}
where we have the period $P_{ph}$ (in s) for a known combination of $B_{ph}$ (in G), $\rho_{ph}$ (in kg m$^{-3}$) and $T_{ph}$ (in MK). A similar analysis on the cases with added thermal conduction can be found in Appendix \ref{sec:appendix}. 

Finally we show in Table \ref{tab:diagnostic}, some examples of using the above formula to calculate the periods of oscillatory reconnection for different combinations of parameters for a flaring coronal plasma. One thing that needs to be stressed is that Equation (\ref{eq:PnoTCb}) has been derived from a set of values that reflects the average conditions in the solar corona. As a result, we need to be cautious when extrapolating the above relation for values outside of that parameter space, as we may end up with non-physical results. However, the derived relation can be a useful plasma diagnostic tool in coronal conditions, and needs to be tested further against observational periodic signals, that could be attributed to oscillatory reconnection. Such periodic signals in the solar atmosphere include, but are not limited to quasi-periodic pulsations (QPPs) of solar \citep[e.g.][]{Kupriyanova2016SoPh} and stellar flares \citep[e.g.][]{2019A&A...629A.147B}, quasi-periodic chromospheric \citep[e.g.][]{DePontieu2011Sci} and coronal jets \citep[e.g.][]{2019ApJ...874..146H,Mandal2022a}, quasi-periodic fast-propagating (QFP) magnetosonic wave from the eruption of a magnetic flux rope \citep[e.g.][]{2018ApJ...853....1S} and periodicities correlated with Type III radio bursts \citep{2021A&A...650A...6C}. A detailed discussion of the different phenomena attributed to oscillatory reconnection has already been presented in the Section \ref{sec:introduction}. The fundamental nature of oscillatory reconnection in perturbed magnetic X-points, indicates that our derived plasma diagnostic tool can be employed to study the periodicities in the different cases of periodic signals attributed to oscillatory reconnection.


To summarize, our series of parameter studies have explored the effects of temperature, density and magnetic field strength on the periodicity of oscillatory reconnection in a hot coronal plasma, expanding the earlier results of \citet{Karampelas2022a}. Our findings show that the period of the oscillation depends on the underlying characteristics of the plasma near the null point. Taking into additional account the independence of the periodicity of oscillatory reconnection from the strength and type of the initial, perturbing pulse \citep{Karampelas2022b}, we have now developed a first quantitative formula to be used as a plasma diagnostic, opening the possibility of using this mechanism within the context of coronal seismology.


\begin{acknowledgments}
All authors acknowledge UK Science and Technology Facilities Council (STFC) support from grant ST/T000384/1. K.K. also acknowledges support by an FWO (Fonds voor Wetenschappelijk Onderzoek – Vlaanderen) postdoctoral fellowship (1273221N). This work used the Oswald High Performance Computing facility operated by Northumbria University (UK). 
\end{acknowledgments}


\appendix
\section{Empirical formula for added thermal conduction} \label{sec:appendix}
In \S\ref{sec:discussions3} we derived an empirical formula (see Equations \ref{eq:PnoTC} and \ref{eq:PnoTCb}) that connects the period of oscillatory reconnection with the characteristic strength of the magnetic field, the background density and the equilibrium plasma temperature. We did this by merging the derived relations of each fit in the data sets without anisotropic conduction, featured in \S\ref{sec:results}. A similar formula can be derived for the period, magnetic field strength and density when we include anisotropic thermal conduction for a plasma temperature of $T_{ph}=1$\,MK:
\begin{equation}\label{eq:PTC}
\frac{P_{ph}}{t_0} = 3.398\frac{B_0}{B_{ph}} + 3.352 \sqrt{\frac{\rho_{ph}}{\rho_0}} - 2.701\pm 0.547
\end{equation}
where we used the period of model 2 (see Table \ref{tab:param}) with thermal conduction switched on ($P_{ph}=(4.049 \pm 0.311)\,t_0$) in order to calculate the penultimate right-hand term. Similarly to Equation (\ref{eq:PnoTC}), we also include the maximum error derived from the different combinations of errors in the values of $P_{ph}$ and the coefficients of the fits.
When written in SI units, except from the magnetic field that is in Gauss, the previous relation takes the form:
\begin{equation}\label{eq:PTCb}
P_{ph} = \frac{38.07}{B_{ph}} + 26.08 \, \times 10^{6} \sqrt{\rho_{ph}} - 21.01\pm 4.26
\end{equation}
where we again take $B_{ph}$ in G and $\rho_{ph}$ in kg m$^{-3}$, for $T_{ph}=1$\,MK.

One drawback of the current study is the fact that, due to numerical reasons, implementing thermal conduction for the setups with high coronal temperatures ($>1$\,MK) lead to very costly and slow to perform simulations for our resolution of choice. That means that Equations (\ref{eq:PTC}) and (\ref{eq:PTCb}) lack the temperature term of Equations (\ref{eq:PnoTC}) and (\ref{eq:PnoTCb}) and can only be valid for plasma with temperatures near 1 MK. However, that might not necessarily hinder our analysis. By comparing the two sets of equations, we see that the respective coefficients for each independent variable are very close in value to each other, when considering either the dimensionless or dimensionalized expressions respectively. Also, from our past studies \citep{Karampelas2022a,Karampelas2022b} and the results of the parameter studies for the magnetic field and density, we know that the addition of thermal conduction only increases the values of the oscillation period by a small amount. Given that for $T=1$\,MK, the parallel to the magnetic field thermal conduction coefficient $\kappa_{\parallel}$ is already many orders of magnitude larger than the perpendicular one $\kappa_{\perp}$, it is unlikely that an increased temperature will significantly change the response of our setups to anisotropic thermal conduction. We thus conclude that our empirical formula without anisotropic thermal conduction (see Equations \ref{eq:PnoTC} and \ref{eq:PnoTCb}) are accurate for solar coronal plasma.

\bibliography{paper}{}
\bibliographystyle{aasjournal}

\end{document}